\documentclass[iop]{emulateapj}
\usepackage{graphicx}
\usepackage{amsmath}  
\usepackage{amssymb} 
\usepackage[lofdepth,lotdepth,caption=false]{subfig}
\usepackage{natbib}
\usepackage{apjfonts}
\usepackage{epsf}
\usepackage{enumitem}

\newcommand{\nstars}{226}   
\newcommand{\nands}{nine}
\newcommand{\snmin}{15}
\newcommand{\feh}{\ensuremath{[\rm{Fe/H}]}}
\newcommand{\fehavg}{\ensuremath{\langle[\rm{Fe/H}]\rangle}}
\newcommand{\alphafe}{\ensuremath{[\alpha/\rm{Fe}]}}
\newcommand{\alphafeavg}{\ensuremath{\langle[\alpha/\rm{Fe}]\rangle}}
\newcommand{\alphafeatm}{\ensuremath{[\alpha/\rm{Fe}]_{\rm{atm}}}}

\newcommand{\Teff}{\ensuremath{T_{\rm{eff}}}} 
\newcommand{\logg}{\ensuremath{\log(g)}}

\def\spose#1{\hbox to 0pt{#1\hss}}

\shorttitle{Alpha Abundances in M\,31 Dwarfs}
\shortauthors{Vargas~et~al.}

\begin{document}

\title{The Distribution of Alpha Elements in Andromeda Dwarf Galaxies}

\author{Luis C. Vargas\altaffilmark{1}} 
\author{Marla C. Geha\altaffilmark{1}}
\author{Erik J. Tollerud\altaffilmark{1,\dag}}

\affil{\altaffilmark{1}Department of Astronomy, Yale University, 260 Whitney Ave., New Haven, CT~06511, USA; luis.vargas@yale.edu}
\altaffiltext{\dag}{Hubble Fellow}

\begin{abstract}

We present alpha to iron abundance ratios for \nstars\, 
individual red giant branch stars in \nands\, dwarf galaxies 
of the Andromeda (M\,31) satellite system.  The abundances 
are measured from the combined signal of Mg, Si, 
Ca, and Ti lines in Keck/DEIMOS medium-resolution spectra. 
This constitutes the first large sample of alpha abundance ratios 
measured in the M\,31 satellite system. The dwarf galaxies in our sample 
exhibit a variety of alpha abundance ratios, with the average values in 
each galaxy ranging from approximately solar (\alphafe$\sim{+0.0}$) 
to alpha-enhanced (\alphafe$\sim{+0.5}$).  These variations do 
not show a correlation with internal kinematics, environment, 
or stellar density.  We confirm radial gradients in the iron 
abundance of two galaxies out  of the five with sufficient data 
(NGC~185 and And~II).  There is only tentative evidence for 
an alpha abundance radial gradient in NGC 185.  We homogeneously 
compare our results to the Milky Way classical dwarf spheroidals, 
finding evidence for wider variation in average alpha abundance.
In the absence of chemical abundances for the M\,31 stellar halo, 
we compare to the Milky Way stellar halo.  A stellar halo comprised 
of disrupted M31 satellites is too metal-rich and inconsistent with 
the Milky Way halo alpha abundance distribution even if considering 
only satellites with predominantly old stellar populations. The M\,31 
satellite population provides a second system in which to study 
chemical abundances of dwarf galaxies and reveals a wider variety 
of abundance patterns than the Milky Way. 

\end{abstract}


\section{Introduction}\label{intro_sec}

\begin{deluxetable*}{lccccccccrr}[t!]
\tabletypesize{\footnotesize}
\tablecolumns{11}
\tablecaption{Observed Properties of M\,31 Satellite Sample\label{table: properties}}
\tablewidth{\textwidth}
\tablehead{
\colhead{Galaxy} & 
\colhead{M$_{V}^{a}$} & 
\colhead{$\sigma_{los}^{b} \rm{[km/s]}$} & 
\colhead{V$_{rms}^{c} \rm{[km/s]}$} & 
\colhead{$\mu_{V}^{d} \rm{[mag/arcsec^{2}]}$} & 
\colhead{r$_{1/2}^{d}$} \rm{[arcmin]} & 
\colhead{PA$^{d}$ [$\rm{^{o}}$]} & 
\colhead{$\epsilon^{d}$} & 
\colhead{$(m-M)_{0}^{e}$} &
\colhead{$N(\rm{[\alpha/Fe]})^{f}$} &
\colhead{$N(\rm{[Fe/H]})^{g}$}
}
\startdata
M 32 & $-16.4\pm 0.2$ & $ 29.9\pm 3.0$ & $ 31.5\pm 2.9$ & $ 11.1\pm 0.0$ & $  0.5\pm 0.1$ & $159.00$ & $ 0.25$ & $ 24.5\pm 0.2$ & $3$ & $32$ \\
NGC 185 & $-14.8\pm 0.1$ & $ 24.0\pm 1.0$ & $ 28.3\pm 2.8$ & $ 20.8\pm 0.0$ & $  1.5\pm 0.0$ & $ 35.00$ & $ 0.15$ & $ 24.0\pm 0.1$ & $71$ & $299$ \\
NGC 147 & $-14.6\pm 0.1$ & $ 16.0\pm 1.0$ & $ 23.3\pm 1.6$ & $ 21.2\pm 0.0$ & $  2.0\pm 0.0$ & $ 25.00$ & $ 0.41$ & $ 24.3\pm 0.1$ & $7$ & $184$ \\
And VII & $-13.3\pm 0.3$ & $ 13.0\pm 1.0$ & $ 13.0\pm 1.0$ & $ 23.2\pm 0.2$ & $  3.5\pm 0.1$ & $ 94.00$ & $ 0.13$ & $ 24.4\pm 0.1$ & $29$ & $90$ \\
And II & $-12.4\pm 0.2$ & $  7.8\pm 1.1$ & $ 13.4\pm 2.1$ & $ 24.5\pm 0.2$ & $  6.2\pm 0.2$ & $ 46.00$ & $ 0.10$ & $ 24.0\pm 0.1$ & $56$ & $248$ \\
And I & $-11.8\pm 1.0$ & $ 10.2\pm 1.9$ & $ 10.2\pm 1.9$ & $ 24.7\pm 0.2$ & $  3.1\pm 0.3$ & $ 22.00$ & $ 0.22$ & $ 24.3\pm 0.1$ & $7$ & $31$ \\
And III & $-10.2\pm 0.3$ & $  9.3\pm 1.4$ & $  9.3\pm 1.4$ & $ 24.8\pm 0.2$ & $  2.2\pm 0.2$ & $136.00$ & $ 0.52$ & $ 24.3\pm 0.1$ & $8$ & $35$ \\
And V & $ -9.6\pm 0.3$ & $ 10.5\pm 1.1$ & $ 10.5\pm 1.1$ & $ 25.3\pm 0.2$ & $  1.4\pm 0.2$ & $ 32.00$ & $ 0.18$ & $ 24.4\pm 0.1$ & $40$ & $80$ \\
And X & $ -7.4\pm 0.1$ & $  6.4\pm 1.4$ & $  6.4\pm 1.4$ & $ 26.3\pm 1.1$ & $  1.3\pm 0.1$ & $ 46.00$ & $ 0.44$ & $ 24.1\pm 0.1$ & $5$ & $16$
\enddata
\renewcommand{\thefootnote}{\alph{footnote}}
\footnotetext[1]{Absolute magnitude (M$_{V}$) sources: And~X \citep{Brasseur2011b}, And~I,III,V,VII \citep{Tollerud2012a}, And~II, NGC~147, NGC~185, M~32 \citep[][and references therein]{McConnachie2012a}.}
\footnotetext[2]{Velocity dispersion sources: And~I,III,V,VII,X \citep{Tollerud2013a}, And~II \citep{Ho2012a}, NGC~147,185 \citep{Geha2012a}, M~32 \citep{Howley2013a}.}
\footnotetext[3]{V$_{rms}$ calculated as quadrature sum of dispersion and rotation velocity; non-zero rotation velocities taken from same sources as velocity dispersions.}
\footnotetext[4]{Surface brightness ($\mu_{V}$), half-light radius (r$_{1/2}$), position angle (PA), and ellipticity ($\epsilon$) from compilation by \citet{McConnachie2012a}.}
\footnotetext[5]{Distance moduli sources: \citep{Conn2012a}, except And~VII \citep{Tollerud2012a}, And~X \citep{Brasseur2011b}, M~32 \citep{McConnachie2012a}.}
\footnotetext[6]{Number of measurements of \alphafe\, excluding $\feh>-0.5$ stars (see discussion in \S~\ref{ssec_analysis}. All measurements are from S/N $\geq{15}$\AA$^{-1}$.}
\footnotetext[7]{Number of measurements of \feh\, used for computing metallicity radial gradients. All measurements are from S/N $\geq{8}$\AA$^{-1}$.}
\renewcommand{\thefootnote}{\arabic{footnote}}
\end{deluxetable*}

\setcounter{footnote}{0}
The alpha to iron abundance ratio correlates with the star 
formation timescale of a stellar population, its star formation 
history, and initial mass function \citep{Tinsley1979a}. The 
vast majority of stellar alpha element abundances 
measured outside of the Milky Way (MW) have 
been limited to the MW's nearby satellite galaxies. 
Spectroscopic studies have shown that the ratio of alpha 
to iron abundances in MW dwarf galaxies decrease gradually 
from super solar (alpha-enhanced) values at low metallicities to solar or 
sub-solar values at higher metallicities\footnote{We will often refer to the alpha to iron abundance 
ratio, \alphafe, simply as the alpha abundance, and to the iron 
abundance, \feh, as the metallicity.} \citep[e.g.,][]{Shetrone2001a,Geisler2005a,Letarte2010a,
Cohen2010a,Kirby2011b}. Based only on the MW dwarf galaxy 
population, the characteristic (or average) stellar abundance 
properties seem to depend on galaxy mass and/or luminosity. 
The majority of the stars in the more massive dwarf galaxies 
have solar/sub-solar alpha abundances \citep[e.g.,][]{Pompeia2008a,
Letarte2010a,McWilliam2013a}. In contrast, less luminous dwarf 
galaxies tend to have a significant fraction of stars with higher 
alpha abundances \citep[e.g., the Sculptor dwarf galaxy, 
see][]{Starkenburg2013a}. The dependence on galaxy properties 
is also exemplified by the correlation between stellar mass and
average stellar metallicity \citep[e.g.][]{Dekel2003a,Grebel2003a,Kirby2013b}.

The alpha abundance patterns of MW dwarf satellites have 
been interpreted as due to long star formation timescales 
\citep[e.g.,][]{Carigi2002a,Lanfranchi2007a}. Type~II supernovae (SNe) 
begin exploding almost immediately after the onset of star formation 
and their ejecta have large yields of alpha elements \citep[][]{Woosley1995a,Chieffi2004a,Nomoto2006a}.
Their high alpha abundance patterns are imprinted in the most metal
poor stars observed in the MW dwarf galaxies \citep[e.g.,][]{Frebel2010c,Cohen2010a}.
In contrast, the onset of Type~Ia SNe is delayed relative to Type~II SNe, 
and these SNe continue to occur long after the onset of star 
formation \cite[e.g.,][]{Maoz2012b,Matteucci2009a}. Type~Ia SNe
release large quantities of iron \citep[e.g,,][]{Iwamoto1999a}, 
thus lowering \alphafe\, in the ISM. This occurs only after significant 
chemical evolution, thus producing the characteristic trend 
towards lower alpha abundances at higher metallicities. Dwarf 
galaxies with large fractions of low alpha stars thus formed the bulk
of their stellar population in timescales comparable to or longer than the 
characteristic timescale for Type~Ia SNe \citep{Matteucci2009a}. 
Indeed, the more massive MW dwarf galaxies (e.g., Fornax and Leo~I) 
are consistent with having a large fraction of intermediate age 
populations \citep[e.g.,][]{Gallart1999a,Hernandez2000a,De-Boer2012a,Del-Pino2013a}. 
In contrast, stellar systems dominated by enhanced alpha abundances 
reflect a large contribution from Type~II SNe. Thus, they 
formed the bulk of their stellar population in bursts with 
sustained high rates of star formation thus softening the 
decrease in alpha abundances due to the Type~Ia contribution 
\citep{Gilmore1991a}. Alternatively, star formation tapered off 
(or ceased completely) in a time shorter than the characteristic 
Type~Ia timescale, due to either internal or environmental processes.

In contrast to the MW dwarf galaxies, our knowledge of the 
chemistry of the dwarf satellites of Andromeda (M\,31) has 
been limited to their overall iron abundance. The
M\,31 dwarf galaxies appear to follow a luminosity-metallicity 
relationship similar to the MW dwarf satellites \citep{Kalirai2010a,Kirby2013b,Collins2013a,Ho2014a}.
This relation appears to be universal amongst Local Group dwarf 
galaxies \citep{Kirby2013b}. Environment often plays a role 
in galaxy evolution, and it is known that the M\,31 system 
presents interesting contrasts to the MW. The M\,31 dwarf 
galaxy population includes more massive dwarf 
spheroidals, and a larger overall number of massive satellites 
\citep{Martin2013a}, with a broad range of kinematic properties 
\citep[][]{Geha2010a,Tollerud2012a,Tollerud2013a,Collins2013a}. 
M\,31 has very likely experienced a more active accretion 
history than the MW as evidenced by the properties of its stellar halo 
\citep[e.g.,][]{Brown2008a,McConnachie2009a,Gilbert2012a}.  
Measurements of alpha abundances in M\,31 dwarf galaxies can 
thus help elucidate the role of environment via a comparison of 
the alpha abundance pattern of its dwarf galaxies, relative to
those of the MW satellites.

Alpha abundances in dwarf galaxies are also relevant 
for understanding hierarchical formation. In $\Lambda$CDM, 
stellar halos are generally believed to have formed from 
accreted dwarf galaxies \citep{Bullock2005a,Bell2008a,
Cooper2010a,Font2011b}. The average alpha abundances of 
MW dwarf galaxies disagree with the abundances characteristic 
of the MW stellar halo \citep{Venn2004a}. \citet{Robertson2005a} 
pointed out that the population of present-day satellites need not 
resemble the properties of already accreted satellites, which were 
likely more massive and had shorter star formation timescales. 
However, the contribution of present-day dwarf galaxies to the
stellar mass of a MW-mass halo may vary stochastically and 
be more important in other systems \citep{Cooper2010a}. Thus,
the chemical inventory in present day dwarf galaxies may be
useful for informing simulations of stellar halo formation. 

In this paper, we present a comprehensive study of the 
alpha abundance of \nands\, dwarf galaxies in the 
M\,31 system, using Keck/DEIMOS medium$-$resolution
spectroscopy of individual red giant branch stars (RGBs). 
These data constitute the first published alpha abundance for 
dwarf galaxies in the M\,31 system. We describe our data 
sample in \S~\ref{sec_data}. We explain our chemical 
abundance technique in \S~\ref{sec_analysis}, focusing in 
particular in those aspects of the analysis relevant to low 
signal to noise data. We present our results in \S~\ref{sec_results}, 
and discuss our results in the context of dwarf galaxy evolution 
and $\Lambda$CDM in \S~\ref{sec_disc}.

\section{Observations and Data Reduction}\label{sec_data}

We measure metallicities and alpha abundances for a 
large sample of stars in \nands\, dwarf satellites of M\,31. 
The sample consists of RGB stellar spectra taken with 
Keck/DEIMOS \citep{Faber2003a} during multiple runs. 
We include data for NGC~147, NGC~185 \citep{Geha2010a}, 
And~II \citep{Ho2012a}, And~VII \citep{Ho2014a}, and
new observations for And~V and And~X (described in \S~\ref{ssec_new}).
The remainder of the spectra were drawn from the Spectroscopic 
and Photometric Landscape of Andromeda's Stellar 
Halo (SPLASH) project, in particular, the dataset presented 
by \citet{Tollerud2012a}. Table~\ref{table: properties} 
lists the dwarf galaxies analyzed, the kinematic, structural 
parameters, and distances assumed throughout this work,
and the number of alpha and iron abundance measurements
per galaxy.

\subsection{Data Reduction}

The full dataset consists of medium-resolution 
(R$\sim{6,000}$) Keck/DEIMOS spectra, taken 
with the 1200~\textit{l}/mm grating and OG550 order 
blocking filter, a central wavelength of \mbox{$\sim{7,500}$ \AA}, 
and either the 0."7 or 1."0 slits. The resulting 
spectral resolution is approximately 1.2 \AA\, (FWHM). 
The spectral setting covers the region from 
\mbox{6300 \AA\, $<\lambda<$ 9100 \AA}, with 
only a small spectral gap between the blue and red chips. 

The data were reduced using the method presented
by \citet{Simon2007a}, who modified the DEIMOS spec2d 
pipeline \citep{Cooper2012a,Newman2013a}
for stellar spectral analysis. The code was used 
to extract one-dimensional, wavelength calibrated
science spectra. The pipeline also outputs an associated sky
spectrum for each slit. Radial velocities were measured by 
cross-correlation with a set of high S/N templates measured 
with the same spectral setup, and the velocities were corrected 
for slit mis-centering. We refer the reader to \citet{Simon2007a}
and \citet{Tollerud2012a} for a more in-depth discussion
of the radial velocity measurement technique.

\subsection{Photometry and Membership Selection}\label{ssec_photometry}

We make use of previously published photometry for our
spectroscopic analysis. The photometry is used in combination
with isochrones to calculate initial temperature (\Teff) estimates, 
and fix surface gravities, \logg, for use in the spectroscopic 
analysis (see \S~\ref{sec_analysis}). We used the $M-T2$ 
photometry for And~I, And~III, And~V, And~VII, and And~X 
from the SPLASH survey (Beaton, \textit{comm}), $RI_{C}$ 
photometry for NGC~147 and NGC~185 \citep{Geha2010a}, 
V-\textit{i} Subaru photometry for And~II \citep{Ho2012a}, and 
Megacam \textit{g'r'i'} photometry for M~32 \citep{Howley2013a}. 
The inhomogeneity in the photometry does not affect the results 
of our analysis.

We rely on the membership selection presented by 
\citet{Tollerud2012a}, \citet{Ho2012a}, \citet{Geha2010a},
and \citet{Ho2014a}. The member stars 
were selected on the basis of radial velocity, position in a 
color-magnitude diagram, distance from the center of the 
dwarf galaxy, and dwarf/giant star spectral diagnostics including 
the strength of the \ion{Na}{1} \mbox{$\lambda8190$} infrared 
doublet. 

The analysis below assumes we have sampled the true
abundance distribution of these galaxies in an unbiased
way.   Given the large luminosity of many of our
galaxies, our spectroscopic member samples are limited
to a small fraction, $<10\%$, of all RGB member stars
down to our limiting magnitude.   Biases in the measured 
metallicity distribution could be injected if stringent color 
cuts were made, since metallicity and color are correlated.  
Our spectroscopic samples sample a wide range of colors, 
as seen in the CMDs shown by \citet{Tollerud2012a}.  
Furthermore, the majority of targets with colors discrepant 
from the visible RGBs are also classified as non-members 
based on their radial velocity.   Another source of bias may be
due to comparing different stellar populations, e.g. RGBs
vs main sequence stars. Most chemical abundance studies
of dwarf galaxies are based on the RGB population, and
it is to these populations that we compare our sample
throughout.   In our analysis, we take into account our 
limited sampling of these galaxies, but caution the reader 
that minority populations in chemical abundance space 
may remain undetected.

\subsection{New And~V and And~X Masks}\label{ssec_new}

We observed two new DEIMOS masks for And~V and one for
And~X during an observing run on September 14-16, 2012,  
supplementing the previous DEIMOS observations of both 
galaxies by \citet{Tollerud2012a}. Objects with membership 
probability \mbox{$P > 0.05$} and S/N $< 20\,$\AA$^{-1}$ 
from \citet{Tollerud2012a} were re-observed to provide 
higher S/N spectra. Each mask was observed between 
1 and 4 hours.

We calculated membership probabilities for these new
spectra using very similar criteria as for the other data. 
The probability based on radial velocity, $P_{vel}$ was 
modeled as a Gaussian centered on the systemic velocity 
determined by \citet{Tollerud2012a}, and with $\sigma$ 
equal to the velocity dispersion measured in that work. 
The positional probability, $P_{pos}$, was set to 1 within
$r_{1/2}$ and modeled as a decreasing exponential 
elsewhere, using the half-light radius given in Table~\ref{table: properties}. 
We also calculated a probability based on position on a 
color-magnitude diagram.  A star was 
given $P_{cmd} = 1$ whenever its color  fell between the 
$\feh = -2.5$ and $\feh = -0.5$ Yale isochrone RGB ridges 
(assuming $\alphafe = +0.4$ and 12 Gyr). If the color was 
blue-ward/red-ward of these color bounds, $P_{cmd}$ was 
down-weighted by the difference between the measured color 
and the blue/red ridge using a Gaussian with $\sigma = \sigma_{color}$. 
Finally, we used the two dwarf-giant star discriminants, the 
\ion{Na}{1}-($V-I$) spectral line diagnostic by \citet{Gilbert2006a} 
and the \ion{Mg}{1}-$\Sigma$CaT diagnostic by \citet{Battaglia2012b}. 
These diagnostics only helped to confirm a few radial velocity 
non-members in our masks, so we did not incorporate them 
into our membership probability. The membership probability 
$P_{memb}$ for each star was then calculated as 
$P_{memb} = P_{vel}\times{}P_{pos}\times{}P_{cmd}$.
We use a low membership threshold of $P_{memb} = 0.05$. 
We note that field contamination at the radial velocity of 
these two galaxies from MW and M\,31 stars is expected to 
be small, and that our samples are concentrated mainly on 
the inner two half-light radii of both galaxies, where the 
contrast between member and non-member stars is highest.

\section{Chemical Abundance Analysis}\label{sec_analysis}

We measure the alpha to iron abundance ratio, \alphafe,
in individual RGB spectra using the spectral synthesis
technique described first by \citet{Kirby2008a},
expanded by \citet{Vargas2013a} (V13). We describe 
the technique in \S~\ref{ssec_preanalysis}-\ref{ssec_analysis}. 
Due to the distance to the M\,31 system, our spectra 
have lower S/N than those analyzed by \citet{Kirby2008a} 
or V13.  We thus perform Monte Carlo tests to assess 
the recoverability of \alphafe, as described in \S~\ref{ssec_analysis_synth}.

\subsection{Photometric Inputs and Synthetic Grid}\label{ssec_preanalysis}

\begin{figure*}[tpb!]
\centering
\includegraphics[width=.98\textwidth]{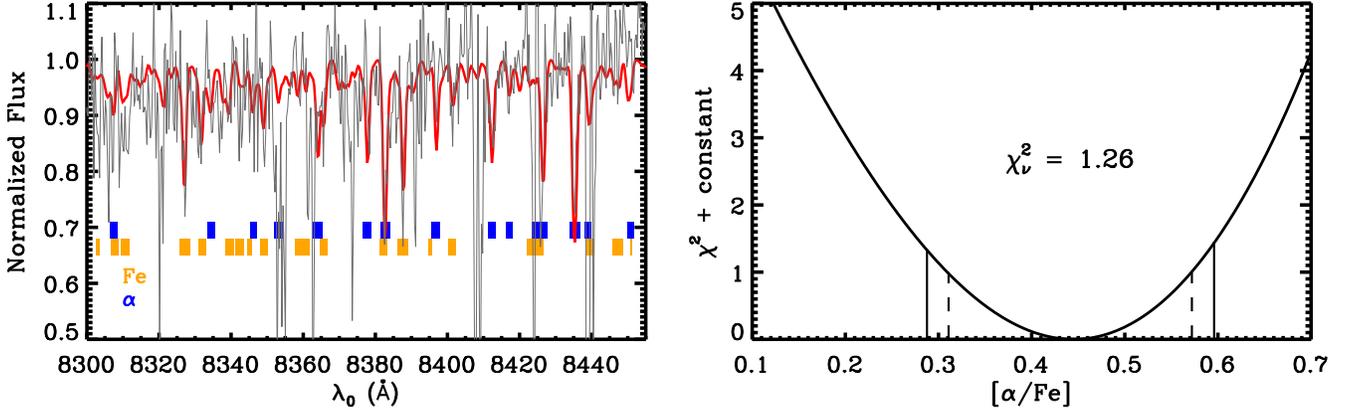}	
\caption{\textit{Left}: Sample Keck/DEIMOS spectrum of a 
red giant branch star in And~VII.  The star has \feh $= -1.22$ and
\alphafeatm $= +0.44$.  Only a small spectral range is shown 
for illustration purposes. Regions used for fitting Fe and alpha
elements are shown in orange and blue, respectively.  \textit{Right:} 
\textit{Un}reduced $\chi^{2}$ as a function of model \alphafeatm. 
The random uncertainty and full uncertainty (including the 
systemic error floor) are shown as dashed and solid lines, 
respectively.  The reduced $\chi^2$ for the best-fitting model, 
$\chi^2_{\nu}$, is shown in the plot. The procedure for 
determining abundances and their uncertainties is described 
in Section~\ref{ssec_analysis}.}
\label{fig: spectrum}
\end{figure*}

Our abundance technique measures iron and alpha 
abundances by finding the best$-$fit between each 
DEIMOS stellar spectrum and a synthetic grid of 
spectra. The grid has a fine sampling in \Teff, surface 
gravity (\logg), stellar metallicity, and \alphafeatm, where 
\alphafeatm\, stands for the abundance ratio of alpha 
elements relative to iron, all varied in lockstep relative 
to their solar-scaled composition at a given metallicity. 
The metallicity represents the lockstep abundance 
variation of all elements other than H and He, relative 
to their solar abundance. In practice, it is measured only 
from Fe lines. The full list of alpha elements is 
enumerated by \citet{Kirby2011c}. In practice, 
only Mg, Si, Ca, and Ti lines are used when 
fitting the spectra. When referring to the grid, we will 
use \alphafeatm, reserving unscripted \alphafe\, for the 
"average" alpha abundance ratio, defined below. 

Prior to the actual fitting, we determine \logg\, 
from photometry. The alternate 
method of measuring \logg\, from the spectra is not 
feasible due to a lack of measurable \ion{Fe}{2} features 
needed for fixing ionization equilibrium. From the 
photometry, we also obtain an initial estimate of \Teff\, 
which is then refined during the spectral fitting. We 
measure the photometric \Teff, and \logg\, from the 
isochrone that passes closest to the reddening and 
extinction-corrected colors and magnitudes listed in
\S~\ref{ssec_photometry}, out of a grid of 
isochrones with varying metallicity from $\feh^{iso} = -2.5$ 
to solar, fixed age (12 Gyr), and $\alphafe^{iso}$(=+0.4). 
We calculate these values independently using the 
Yale$-$Yonsei \citep{Kim2002a}, Dartmouth \citep{Dotter2008a}, 
and PARSEC \citep{Bressan2012a} isochrone libraries. 
$\alphafe^{iso} = +0.4$ except for PARSEC, for which 
we use $\alphafe^{iso} = +0.0$ due to the present lack 
of $\alphafe^{iso} = +0.4$ isochrones. We measure 
the uncertainties in $\Teff$ and $\logg$ from 500 Monte Carlo 
resamplings of each star's color and magnitude, and using the standard 
deviation of the distribution of parameters. We average the 
results from the three independent isochrone libraries (using 
variance-weighted means).

Since our RGB sample may contain a wide spread of 
ages and alpha enhancements, we check whether using
a different fixed age and $\alphafe^{iso}$ results in significant 
systematic shifts in \Teff\, and \logg, which would then
influence the abundance analysis. We test for the effect 
of changing $\alphafe^{iso}$ and age separately, using the 
Yale isochrone grid and recalculating \Teff\, and \logg, for 
the stars in NGC~147, keeping the other parameters as before. 
If using $\alphafe^{iso}=+0.0$, \logg\, shifts by only $-0.01$ dex, 
whereas the initial estimates of \Teff\, shift by $-15$ K. If instead 
we use an age of 4 Gyr, \logg\, changes by \mbox{0.16 dex}, 
and the estimates of \Teff\, by \mbox{$-56$ K}. The effect
of these systemic shifts, in particular \logg\, which is not 
refined by the spectral analysis, will translate to systemic 
shifts in the abundances smaller than 0.05 dex, smaller
the error floors in our measurements (see next subsection). 

\subsection{Abundance Analysis}\label{ssec_analysis}

After determining \logg\, and the initial estimate of 
the \Teff\, from isochrones, we measure the abundances 
using the same set of steps as V13. Prior to fitting, the 
science spectra are normalized, and the spectral resolution 
is refined as a function of wavelength. We use fits to sky 
emission lines extracted for each slit to measure the 
variation in resolution with wavelength, $\sigma_{sky}(\lambda)$. 
We adjust $\sigma_{sky}$ with a constant rescaling factor 
for each spectral mask, which encodes the difference in slit 
imaging between the sky lines and the science spectra, i.e. 
$\sigma(\lambda) = k\times\sigma_{sky}(\lambda)$.
The factor $k$ is allowed to vary in a first pass of the
abundance code (described below), and then fixed for 
each mask as the average $k$ for all stars in the mask. 
As in V13, we only fit spectral regions sensitive to 
Fe or alpha elements, excluding regions with significant 
telluric contamination (same regions as defined by \citealt{Kirby2009a}) and 
a few strong lines such as the calcium triplet, which are not 
modeled adequately in the synthetic spectra. In addition, 
we now mask regions affected by strong sky emission. 
While sky lines are nominally subtracted by the pipeline, 
the subtraction is not perfect, an effect that is more 
important at low S/N. We measure the median sky flux 
in 500 \AA$-$wide non$-$overlapping segments. We 
identify pixels where $f_{sky}(\lambda) > f_{sky,median}$, 
and $f_{sky}(\lambda) > 0.25 \times f_{star}(\lambda)$, and 
mask them out when fitting the spectra. 

We next fit each normalized spectrum against the 
aforementioned synthetic grid to find the best-fitting
\feh\, and \alphafeatm, while also allowing \Teff\, to vary 
from the initial photometric estimate. The grid's \feh\, and 
\alphafeatm\, ranges are $-4.8\leq\feh\leq0.0$ and $-0.8\leq\alphafeatm\leq+1.2$, 
respectively, with a gradation of 0.1 dex in both parameters. 
In practice, \Teff\, and \feh\, are fit simultaneously, while
\alphafeatm\, is fit in a separate step. These steps are iterated
to obtain convergence in all three parameters. During the 
fitting, we linearly interpolate the fluxes between the nearest 
spectra at each pixel wavelength. We only fit those spectral 
regions sensitive to changes in Fe or alpha element abundance 
(where alpha is one of Mg, Si, Ca, or Ti), and minimize the 
pixel$-$by$-$pixel flux variation between the synthetic models 
and the spectra using the Levenberg$-$Marquard minimization 
code \textit{mpfit}. 

The uncertainty in \feh\, includes the covariance between 
\feh\, and \Teff, resulting in a larger error than if either parameter 
was fit independently.  For \alphafeatm, V13 found that a 
significant fraction of $\chi^{2}$ contours in the $\chi^{2}$
minimization were asymmetric, with a tendency for the 
$\chi^{2}$ contours flattening out towards lower \alphafeatm.  
Thus, they reported asymmetric \alphafeatm\, error bars by 
finding the values of \alphafeatm\, that satisfy $\chi^2$ = $\chi^2_{min}+1$, 
where $\chi^2_{min}$ is the $\chi^{2}$ value for the best-fitting model.  
We adopt that procedure here and report both lower and 
upper \alphafeatm\, uncertainties, $\sigma^{-}$ and $\sigma^{+}$. 
We note that in the few cases where the $\chi^{2}$ contour 
does not rise to $\chi^2_{min}+1$, we adopt a lower 
uncertainty in \alphafeatm\, equal to the entire range 
in \alphafeatm\, between the best-fit value and the
lower edge of the grid \mbox{(\alphafeatm\,$= -0.8$)}.
Finally, we add in quadrature an uncertainty floor to
all \feh\, ($\sigma=0.113$) and \alphafeatm\, ($\sigma=0.082$) 
measurements as in V13. The uncertainty floor reflects the 
difference between DEIMOS and literature high-resolution 
abundance measurements in the limit of high S/N data (i.e., 
in the regime of small internal uncertainties).  To illustrate 
the technique, we show in Figure~\ref{fig: spectrum} (left panel) 
a portion of a DEIMOS spectrum of an RGB in And~VII, 
together with the best-fitting synthetic spectrum. The right 
panel shows the $\chi^{2}$ fit for \alphafeatm, as well as the 
uncertainties measured as described above.

\begin{figure}[tpb!]
\centering
\includegraphics[width=.48\textwidth]{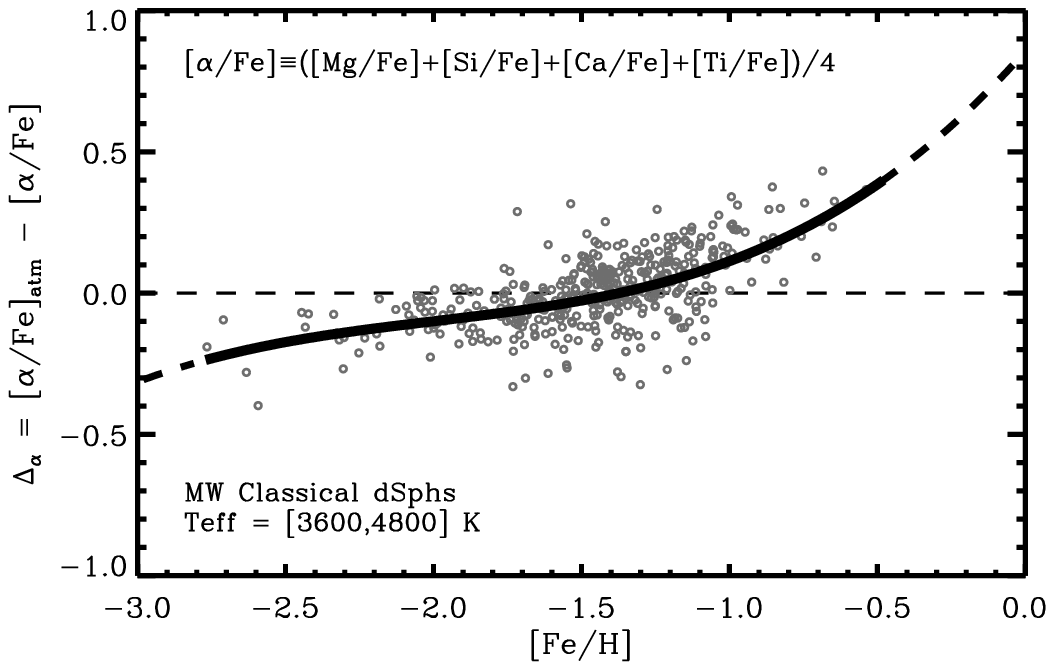}	
\caption{Difference between \alphafeatm\, and the arithmetic mean 
of the abundance ratios: [Mg/Fe], [Si/Fe], [Ca/Fe], and [Ti/Fe], plotted 
as a function of iron abundance, for RGB stars in six MW dSphs in 
the same range of \Teff\, as our program stars. The mean is denoted 
as \alphafe\, throughout the paper.  The solid line is a third-order polynomial 
fit to the difference $\Delta\equiv\alphafeatm-\alphafe$, between $-2.8<\feh<-0.5$, 
using the comparison MW dSph stars. We apply this correction to \alphafeatm\, 
to calculate \alphafe\, for our sample. See \S~\ref{ssec_analysis} for more 
details. The dotted extensions are an extrapolation based on the polynomial
describing the solid line.  Unless stated otherwise, we do not consider the \alphafe\,
values in the extrapolated region in our analysis.}
\label{fig: alphaatm_alphafeavg}
\end{figure} 

\begin{figure*}[t!]
\centering
\includegraphics[width=.8\textwidth]{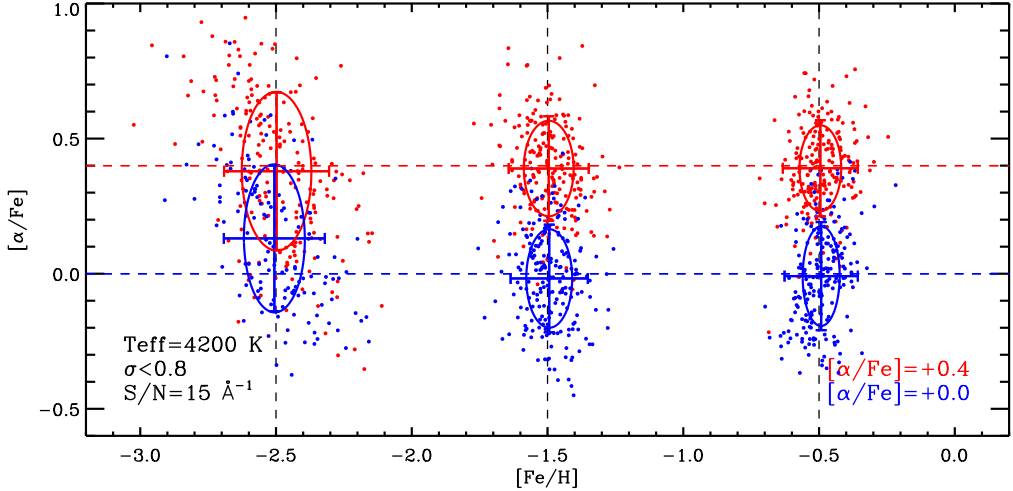}	
\caption{Distribution of recovered \alphafe\, and \feh\, measurements 
at low S/N using mock synthetic spectra.  We created mock spectra
for the following parameter combinations: \feh\, = [-2.5,-1.5,-0.5],
and \alphafe\, = [+0.0,+0.4]. For each parameter set, we draw from 
our synthetic spectral library (degraded to DEIMOS resolution 
wavelength coverage, and desired S/N) to create  200 mock 
spectra, which we then process through our abundance pipeline. 
The ellipses show the $1\sigma$ uncertainties.  We thus can distinguish 
between an "alpha-enriched" (+0.4) population and a "alpha-solar" 
(+0.0) population at this low S/N, although the latter will be slightly 
biased at $\feh<-2.5$.  We restrict our \alphafe\, measurements to 
S/N$>$15\AA$^{-1}$ spectra.}
\label{fig: alpha_synthetic}
\end{figure*} 

We report the alpha abundance ratio, \alphafe, which
we define as the arithmetic mean of  [Mg/Fe], [Si/Fe], 
[Ca/Fe], and [Ti/Fe] \footnote{Throughout the paper, 
we will simply refer to \alphafe\, as the alpha abundance 
for a single star, reserving the bracket notation $\langle...\rangle$ for 
\textit{averages over various population bins}.}. 
We measure \alphafe\, from the \alphafeatm\, parameter 
fit by the abundance code, because the individual alpha
elements are not measurable for most stars due to our 
low S/N. We calculate a correction term for converting 
\alphafeatm\, to \alphafe, using DEIMOS spectra of Milky 
Way classical dSphs originally presented by \citet{Kirby2011b}. 
We have reanalyzed this sample in order to ensure
maximum homogeneity between the Milky Way and M\,31
samples. The Milky Way dSph spectra have higher S/N than 
our program sample, so that \alphafe\, can be measured 
directly from the arithmetic mean of [Mg/Fe], [Si/Fe], [Ca/Fe] 
and [Ti/Fe] for a significant subset of spectra. We fit a third 
order polynomial to the difference between \alphafeatm\, 
and \alphafe\, as a function of \feh, further restricting the MW 
dSph sample only to stars in the \Teff\, range of our M\,31 
dwarf galaxy sample, and for which all four abundance ratios 
were measured. 

Figure~\ref{fig: alphaatm_alphafeavg} shows the difference 
\alphafeatm$-$\alphafe. The relation between \alphafeatm\,
and \alphafe\, varies with \feh\, at least in our \Teff\, range. 
We note that \alphafeatm, which is fit from \textit{all} lines
of Mg, Si, Ca, and Ti, need not be equal to \alphafe, since
the relative strength and number of different spectral lines
and different elements varies with \Teff\, and metallicity.  We
thus favor the definition of \alphafe\, which allows easier 
comparison to other literature results.  We also note that 
without the correction term, the \alphafe\, measurements 
for the Fornax dSph would appear significantly enhanced, 
in contrast to what is measured from the individual alpha 
elements by our reanalysis of MW dSph data, by \citet{Kirby2011b}, 
and by \citet{Letarte2010a}, the latter based on \mbox{R$\sim{20,000}$} 
spectra.  

To determine \alphafe, we apply the correction function discussed 
above (black line in Figure~\ref{fig: alphaatm_alphafeavg}) to all 
\alphafeatm\, measurements, including those in the comparison 
Milky Way dSph sample.  We verified that the vertical scatter 
about the best-fit line is consistent with the data uncertainty.
We thus keep $\sigma_{\alphafe} = \sigma_{\alphafeatm}$.
Due to a lack of calibrators with \mbox{$\feh > -0.5$}, the 
correction term for this metallicity range is only an extrapolation.  
Hence, we restrict our analysis to those stars with $\feh<-0.5$, 
except where noted.  Using a more metal-poor sample, V13 determined a 
positive correction of +0.063 was needed for \alphafeatm\, to equal 
\alphafe. However, this comparison was done on the basis of only 
metal-poor stars with a higher average \Teff.  Using the
Milky Way dSphs sample, we confirm that the difference between
\alphafeatm\, and \alphafe\, is smaller for higher \Teff\, stars.  The 
\feh\, and \alphafe\, measurements are converted from the 
\citet{Anders1989a} abundance scale to the newer \citet{Asplund2009a} 
scale, as in V13. 

\subsection{Precision of \alphafe\, at low S/N: Synthetic Spectral Tests}\label{ssec_analysis_synth}

The distance to the M\,31 dwarf galaxies limits the 
S/N obtainable with reasonable observing times. The 
maximum S/N in our sample is only 25 \AA$^{-1}$. 
In contrast, the sample from which we derived the
correction between \alphafeatm\, and \alphafe\, have
a median S/N between 42 \AA$^{-1}$ (Fornax) and 
150 \AA$^{-1}$ (Draco). While $\chi^{2}$ provides 
an easily measurable estimate of the uncertainty, it is 
plausible to suspect there is a bias in the measurement 
of \alphafe\, where lower \alphafe\, values would be 
harder to observe due to the lower S/N of the line at a 
fixed spectral S/N. To assess the precision of our analysis 
pipeline, we turn to Monte Carlo simulations. 

We create continuum-normalized mock spectra using the 
synthetic spectral library from \citet{Kirby2011c}. We 
convolve the spectra with a Gaussian kernel with FWHM 
$=$ 1.2 \AA, so as to closely reproduce the DEIMOS 
spectral resolution, and bin the degraded spectrum to 
the DEIMOS pixel size and spectral range. To roughly 
simulate the variation of S/N with wavelength, we add
a pseudo$-$continuum derived from a spline fit to a 
DEIMOS template of a cool RGB star. The mock spectra 
have \Teff$=$4,200 K, metallicities ranging from 
[Fe/H] = -4.0 to [Fe/H] = -0.5 (in steps of 0.5 dex)
and \alphafe$=[+0.0,+0.4]$. The \Teff\, value represents
the average \Teff\, in our sample. For each parameter 
combination, $\logg$ value is set by comparison to a grid 
of Dartmouth isochrones, and finding the $\logg$ corresponding 
to the appropriate \Teff, [Fe/H], and \alphafe. For 
$\rm{[Fe/H]} \leq -2.5$, we use the $\rm{[Fe/H]} = -2.5$ isochrones, 
noting that the differences in isochrone RGBs decrease towards lower 
metallicities. We then add Gaussian noise to each pixel consistent
with its flux uncertainty (updating the variance array) such that 
the average S/N in the CaT region is one of $\rm{S/N} = [8, 15, 30]$ \AA$^{-1}$. 
We create 200 mock realizations for each set of parameters at 
each  S/N, and analyze the spectrum with our abundance 
pipeline, using as initial inputs the true \Teff\, and \logg\, of 
the synthetic spectrum. 

Figure~\ref{fig: alpha_synthetic} shows the scatter in \alphafe\, 
against [Fe/H] for the case of a \Teff\, $=$ 4200 K, \mbox{[Fe/H] $= [-2.5,-1.5,-0.5]$}, 
\mbox{\alphafe $= [+0.0,+0.4]$} spectra, showing that for S/N $\gtrsim{\snmin}$ \AA$^{-1}$, 
we can distinguish between a population of \alphafe$=$+0.0 and \alphafe$=$+0.4 
objects. At S/N $<$ \snmin\, \AA$^{-1}$, the MC tests 
suggest we can still recover the average \alphafe\, of a 
population, albeit with a significant increase in the random 
scatter. Because of this, we choose S/N $=$ \snmin\,\AA$^{-1}$ as our 
lower S/N cut for \alphafe. The same set of MC tests show 
that [Fe/H] can be recovered accurately for \mbox{S/N $>$ 8 \AA$^{-1}$}. 
For all analyses involving \alphafe\, we apply the threshold 
\mbox{S/N $>$ \snmin\,\AA$^{-1}$}; we use the lower threshold
for [Fe/H] (\mbox{S/N $>$ 8 \AA$^{-1}$}) only to calculate 
average metallicities and radial metallicity trends. 
\section{Results}\label{sec_results}

\begin{figure*}[tpb!]
\centering
\includegraphics[width=\textwidth]{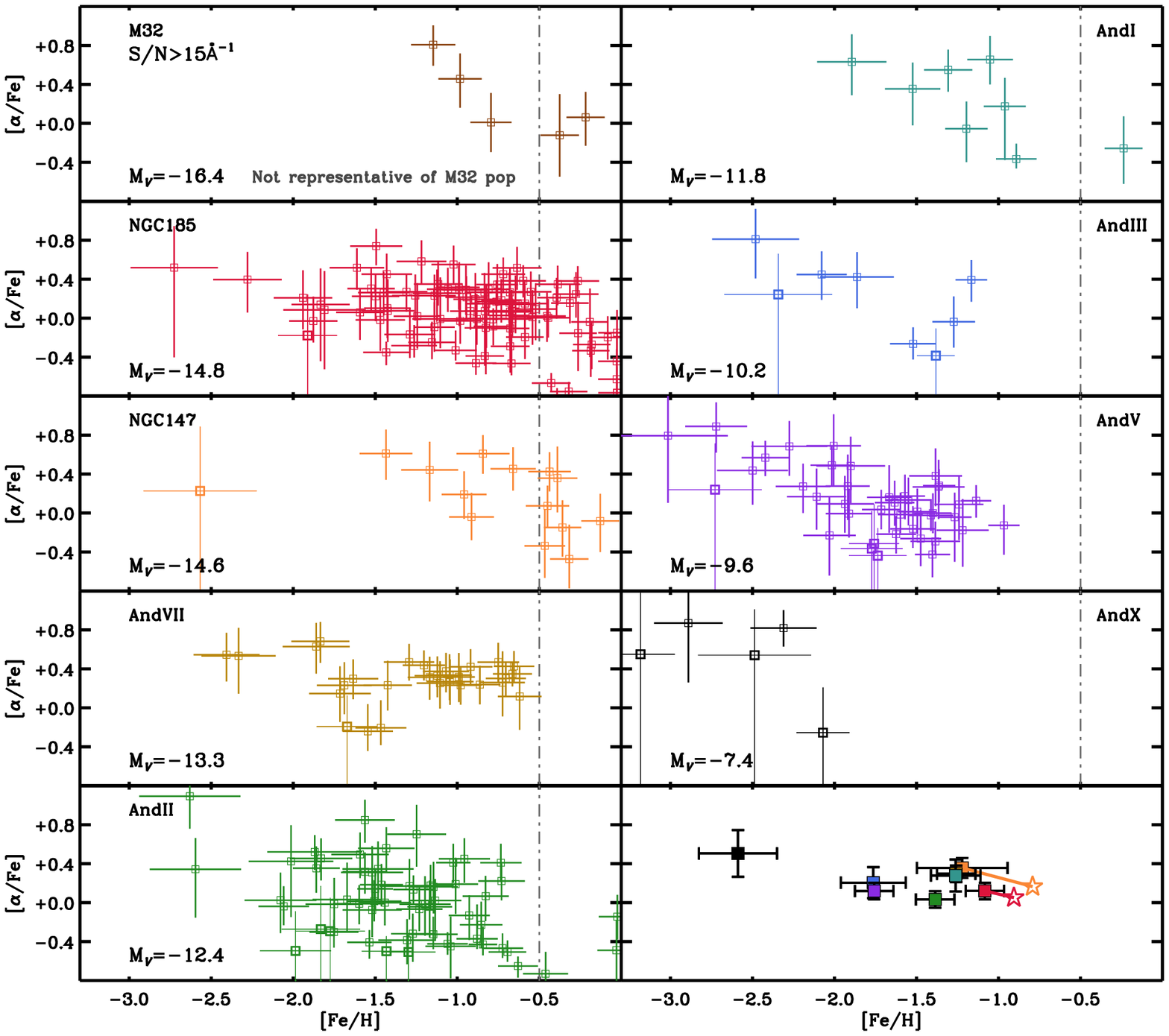}	
\caption{$\alphafe$\, as a function of \feh\, for our entire dataset. 
The dwarf galaxies are shown from most to least luminous in V-band 
(top to bottom). The same color scheme is used to denote each 
galaxy in all following plots. The dash-dotted line at $\feh=-0.5$ indicates 
that \alphafe\, values for higher metallicities are uncertain, due to the 
lack of calibrators to test the relation between \alphafeatm\, and \alphafeavg\, 
(see \S~\ref{ssec_analysis} for discussion).  Each square is a measurement 
of \alphafe\, using only spectra with \mbox{S/N $\geq$ 15 \AA$^{-1}$}. 
\textit{Bottom-right panel}:  We plot the average alpha abundance, \alphafeavg,
using $\feh < -0.5$ stars against the mean metallicity, \fehavg, calculated from
the same stars. For NGC~185 and NGC~147, we include a star symbol
indicating the change in \alphafeavg\, and \fehavg\, if the \feh$>-0.5$ stars 
are included.  We do not include M\,32 in this panel because the \alphafe\,
sample is not representative of the galaxy's population (\S~\ref{ssec_meantrends}).}
\label{fig: alpha_grid}
\end{figure*}

The \alphafe\, abundance ratio is an indicator of the star formation 
timescale in a galaxy. Qualitatively, high average alpha abundances can be 
explained as a being due to short star formation timescales, whereas 
low average alpha abundances are associated with slower star formation 
timescales.  We have measured both alpha and iron abundances in \nands\, 
dwarf galaxies of M\,31, sampling a large range in V-band absolute magnitude 
from M$_V\sim{-7.4}$ (And\,X) to M$_{V}\sim{-16.4}$ (M\,32). 
Figure~\ref{fig: alpha_grid} presents our full dataset, plotted in the 
\alphafe-\feh\, abundance plane. 

\begin{deluxetable}{lrrccrr}[t!]
\tablecolumns{6}
\tabletypesize{\footnotesize}
\tablecaption{Iron and Alpha Abundance Results}
\tablewidth{0pt}
\tablehead{
\colhead{Galaxy} & \colhead{R.A.} & \colhead{Decl.} & \colhead{\feh} & \colhead{\alphafe}  & \colhead{Flag} \\ 
\colhead{} & \colhead{(J2000)} & \colhead{(J2000)} & \colhead{} & \colhead{} & \colhead{} 
}
\startdata
M 32    & 00:42:39.32 & +40:53:27.0 & $-0.79\pm0.13$ & $+0.01^{+0.30}_{-0.31}$ & \\[+2pt]
M 32    & 00:42:42.30 & +40:54:39.6 & $-1.15\pm0.13$ & $+0.81^{+0.20}_{-0.22}$ & \\[+2pt]
M 32    & 00:42:53.34 & +40:51:35.6 & $-0.98\pm0.13$ & $+0.46^{+0.26}_{-0.30}$ & \\[+2pt]
NGC 185 & 00:38:09.18 & +48:13:45.6 & $-1.61\pm0.17$ & $+0.52^{+0.20}_{-0.22}$ & \\[+2pt]
NGC 185 & 00:38:12.51 & +48:13:09.2 & $-1.91\pm0.18$ & $-0.18^{+0.42}_{-0.62}$ & u\\[+2pt]
NGC 185 & 00:38:22.49 & +48:09:47.6 & $-1.49\pm0.16$ & $+0.74^{+0.18}_{-0.20}$ & \\[+2pt]
NGC 185 & 00:38:22.88 & +48:15:47.2 & $-0.99\pm0.13$ & $+0.32^{+0.22}_{-0.23}$ & \\[+2pt]
NGC 185 & 00:38:23.46 & +48:11:07.9 & $-0.83\pm0.13$ & $-0.10^{+0.31}_{-0.48}$ & \\[+2pt]
NGC 185 & 00:38:24.95 & +48:21:54.3 & $-0.82\pm0.13$ & $+0.15^{+0.20}_{-0.20}$ & \\[+2pt]
NGC 185 & 00:38:25.10 & +48:12:52.0 & $-0.68\pm0.12$ & $-0.10^{+0.23}_{-0.23}$ & \\[+2pt]
... & ... & ... & ... & ...
\tablecomments{Table~\ref{table: abundances}
is published in its entirety in the electronic edition
of the Astrophysical Journal. A portion is shown here
for guidance regarding its form and content. Flags on
last column: (1) u - Lower limit uncertainty in \alphafe\, is outside of our grid,
thus we set $\sigma^{-}_{\alphafe}$ = \alphafe - min(\alphafe$_{grid}$) = \alphafe\,$-(-0.8)$.
(2) mr = \feh$>-0.5$ star. (3) f : Only [Fe/H] is available for this star.}
\label{table: abundances}
\end{deluxetable}

A visual inspection of the data shows a variety of abundance
patterns in \mbox{\alphafe-\feh} space between the different 
dwarf galaxies. The plot also indicates the presence of 
significant iron abundance spreads in all dwarf galaxies in 
the sample, confirming previous results based on photometric 
metallicities \citep[e.g.,][]{Kalirai2010a}, and measured directly 
from the calcium triplet \citep{Ho2014a}.
This is the first time that iron abundance spreads are measured 
directly from iron lines in M\,31 satellite galaxies. The sequence
of panels also illustrates qualitatively the correlation between 
luminosity and average iron abundance \citep{Kirby2013b}.  
NGC~147 appears to scatter above the relation. Interestingly, 
this galaxy has been shown to have tidal features in PanDAS 
photometry \citep{McConnachie2009a}. It is thus possible
it was more luminous in the past. If so, this could explain its
relatively high metallicity given its present day luminosity.

\subsection{Mean Chemical Abundance Trends}\label{ssec_meantrends}

Average chemical trends, e.g., the luminosity-metallicity relation, 
are useful in constraining the properties of galaxy evolution as a 
function of luminosity or other global parameters. We thus 
quantify the average (characteristic) alpha and iron chemical 
properties, \alphafeavg\, and \fehavg, for each dwarf galaxy.  We 
calculate \alphafeavg\,using spectra with \mbox{S/N $> 15$ \AA$^{-1}$} 
and $\feh < -0.5$.  We exclude the stars with $\feh > -0.5$ because we 
do not have a calibrating sample that allows us to verify the relation 
between \alphafeatm\, and \alphafe\,(see \S~\ref{ssec_analysis}), but
discuss the effect of excluding these stars in particular galaxies.

We calculate the mean iron abundance, \fehavg, using all spectra 
with \mbox{S/N $> 8$ \AA$^{-1}$}.  We use this lower S/N cut 
because the number of iron-sensitive spectral regions is larger than 
for \alphafeatm, so that even lower S/N are useful for iron abundance 
measurements.  Using this larger sample serves a two-fold purpose. 
It allows us to provide a better constraint on \fehavg\, and to check
whether small number statistics affect our \alphafe\, samples. To do 
the latter, we compare the \feh\, distribution for the alpha sample 
to the \feh\, distribution for the full \feh\, sample for each galaxy.  
With the exception of M\,32 the distributions agree. For M\,32, the 
five stars with \alphafe\, are clustered at $\feh>-1.5$ whereas the \feh\, 
measurements span $-3<\feh<0$.  This suggests that the average 
\alphafeavg\, for the M\,32 sample is not representative of this galaxy's 
alpha enhancement, and do not consider it further in this subsection
\footnote{We do use the M\,32's  data later when comparing \alphafeavg\, 
at fixed metallicity.}.  

We assess the uncertainty in \fehavg\, and \alphafeavg\, 
from the dispersion in the data using the Student $t$-distribution, thus 
taking into account the small sample sizes in various of our galaxies. 
We measure 68\% confidence intervals for both \fehavg\, and \alphafeavg\, 
dividing the confidence interval for a single measurement by $N^{1/2}$ 
(standard error of the mean). In addition, we set an uncertainty floor 
on the means equal to the uncertainty floors described in \S~\ref{ssec_analysis}.
We tabulate our measurements of \alphafeavg\, and \fehavg\,  in 
Table~\ref{table: averages}.  

\begin{figure*}[tpb!]
\centering
\includegraphics[width=\textwidth]{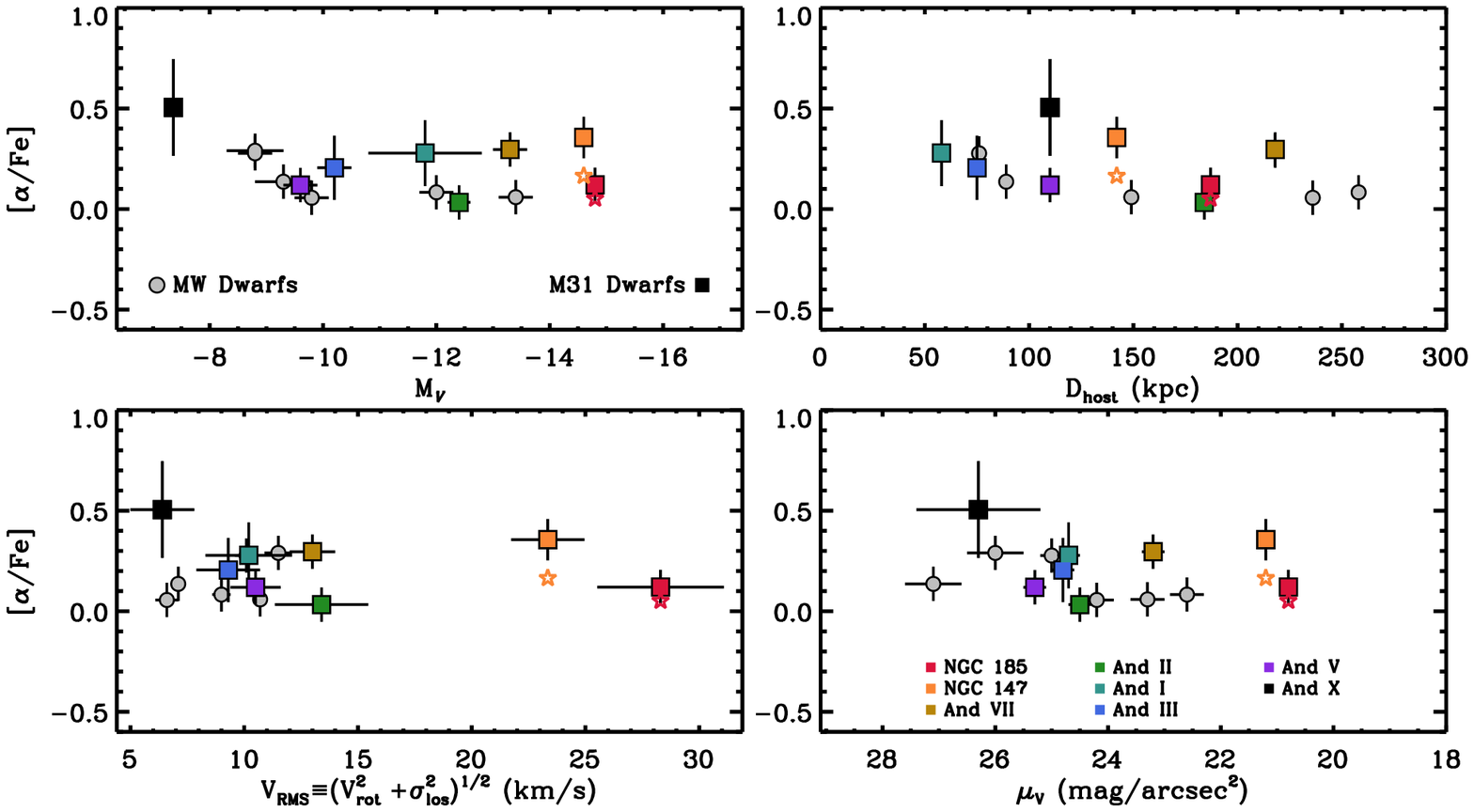}	
\caption{Mean alpha abundance, \alphafeavg, in each of the nine dwarf galaxies as a function 
of (a) luminosity, (b) distance from host galaxy (either M\,31 or Milky Way) 
(c) V$_{rms}$, the quadrature sum of line-of-sight velocity dispersion and rotation, 
and (d) central surface brightness. As discussed in \S~\ref{ssec_meantrends}, we 
only use stars with $\feh<-0.5$. For NGC~147 and NGC~185, we plot a
star symbol for the average alpha abundance calculated when including $\feh>-0.5$ 
stars. The change is non-negligible only for NGC~147. The $1\,\sigma$ error 
in \alphafeavg\, takes into account the variation in sample size 
for each galaxy.  The x-axis values and $1\,\sigma$ uncertainties for each of the 
four properties plotted are taken from the literature, and summarized in 
Table~\ref{table: properties}.  The M\,31 dwarf galaxies are color-coded as 
shown in the bottom right panel (same as in Figure~\ref{fig: alpha_grid}). 
We compare the sample against DEIMOS data for six classical MW dSphs 
analyzed homogeneously (\S~\ref{ssec_analysis}), shown as gray circles.}
\label{fig: alphafe_properties}
\end{figure*}

The bottom-right panel of Figure~\ref{fig: alpha_grid} shows \alphafeavg\, 
versus \fehavg. For this panel only, we have plotted \fehavg\, as measured 
from S/N $> 15$ \AA$^{-1}$, so that both \alphafeavg\, and \fehavg\,
are measured from the same sample. If the \mbox{$\feh>-0.5$} stars 
are included, \alphafeavg\, decreases by $-0.19$ dex (NGC~147) 
and $-0.07$ dex (NGC~185), as shown by the star symbols.
Given \alphafeavg\ $= + 0.36\pm{0.11}$ and $\alphafeavg = +0.12\pm{0.09}$ 
for NGC~147 and NGC~185, respectively, the changes resulting from 
ignoring the \mbox{$\feh>-0.5$} stars cause some tension in \alphafeavg\, 
only in the case of NGC~147. 

The bright dwarf galaxies NGC~185, and NGC~147 appear to have 
a small fraction of stars with \mbox{$\feh>-0.5$}. Since our synthetic 
spectral grid does not extend to \mbox{$\feh>0$}, observed stars at those 
metallicities will be best-fit by a \mbox{$\feh=0$} spectrum. The number of 
such stars is low, ranging from 13/312 in NGC~185 to 31/205 in 
NGC~147.  Based on the results of a CaT study of the same DEIMOS 
sample by \citet{Ho2014a}, excluding these stars would modify \fehavg\, by 
less than 0.15 dex. 

We plot in Figure~\ref{fig: alphafe_properties} the resulting average 
alpha abundance for each galaxy against galaxy luminosity,  distance 
from M\,31, global kinematics, and surface brightness.  We also 
include data for six MW dwarf galaxies (Fornax, Leo~I, Leo~II, 
Draco, Sextans, and Ursa Minor) reanalyzed with our code for 
purposes of comparison and reference (see \S~\ref{ssec_analysis}).
We caution, however, that the homogeneous sample of MW 
satellites is not complete in our magnitude range. Thus we 
avoid comparing the MW and M\,31 satellites 
\textit{as populations}. We defer such comparison to 
\S~\ref{ssec_trend_feh} after adding other bright MW 
satellites from literature data, not included here as 
their sampling may not be appropriate for calculating 
galaxy-wide average chemical properties.

\subsubsection{\alphafeavg\,\lowercase{vs} L\lowercase{uminosity}} 
We compare the chemical properties as a function 
of galaxy luminosity (Figure~\ref{fig: alphafe_properties}, 
top-left panel). The average alpha abundance ratios of M\,31 
dwarf galaxies fainter than \mbox{M$_{V} = -10$} (i.e., And~X, And~III, 
and And~V) are indistinguishable from the Milky Way dwarf 
galaxies of comparable luminosity (Leo~II, Draco, Sextans and 
Ursa Minor). In particular, And~V and And~III have intermediate 
\alphafeavg\, values, consistent with those of Leo~II and Sextans at 
almost the same luminosity.  At higher galaxy luminosities, there is 
a clear distinction between the alpha enhancements of
And~VII ($\alphafeavg\sim+0.3$) and And~II ($\alphafeavg\sim+0.0$). 
At similar luminosities, both MW dwarf galaxies in our 
comparison sample (Fornax and Leo~I) have \alphafeavg\, 
close to And~II (between $\sim{+0.0}$ and +0.1).  The 
MW and M\,31 samples partially overlap in their luminosity range; 
however, the M\,31 sample contains more luminous dwarf 
galaxies. Out of the three brightest galaxies in our sample,  we 
again find a range in alpha abundances, with NGC~147 having 
higher \alphafeavg\, than NGC~185 by $\sim{0.3}$ (although the
value is indistinguishable from NGC~147 if the $\feh>-0.5$ stars 
are included).  In summary, we find that the M\,31 dwarf satellites 
have a range of average alpha abundances enhancements, and there 
is no evidence for this trend being correlated with galaxy 
luminosity, at least for M$_{V}<-9$.  

\begin{deluxetable*}{lrrrrrrrrrr}[t!]
\tablecolumns{11}
\tabletypesize{\footnotesize}
\tablecaption{M31 Dwarf Galaxy Average Chemical Properties}
\tablewidth{0pt}
\tablehead{
\colhead{Galaxy} & \colhead{\fehavg$^{a}$} & \colhead{N$_{\*}$} & \colhead{\alphafeavg$^{b}$} & \colhead{N$_{\*}$} & 
\colhead{\alphafeavg$^{b}$} & \colhead{N$_{\*}$} & \colhead{\alphafeavg$^{b}$} &
\colhead{N$_{\*}$} & \colhead{\alphafeavg$^{b}$} & \colhead{N$_{\*}$} \\ 
\colhead{} & \colhead{} & \colhead{} & 
\multicolumn{2}{c}{[Fe/H] $< -0.5$} & 
\multicolumn{2}{c}{[Fe/H] $= [-2.5,-1.5] $} & 
\multicolumn{2}{c}{[Fe/H] $= [-2.0,-1.0] $} & 
\multicolumn{2}{c}{[Fe/H] $= [-1.5,-0.5] $}
}
\startdata
M 32 & $-1.153\pm0.165$ & $ 32$$^{c}$ & $ 0.425\pm0.319$ & $ 3$$^{d}$ &  &  &  &  & $ 0.425\pm 0.319$ & $ 3$ \\
NGC 185 & $-0.917\pm0.117$ & $299$$^{c}$ & $ 0.120\pm0.085$ & $71$ & $ 0.167\pm 0.085$ & $ 9$ & $ 0.137\pm 0.085$ & $31$ & $ 0.107\pm 0.085$ & $61$ \\
NGC 147 & $-0.491\pm0.117$ & $184$$^{c}$ & $ 0.356\pm0.103$ & $ 7$ &  &  &  &  & $ 0.378\pm 0.121$ & $ 6$ \\
And VII & $-1.240\pm0.117$ & $ 90$ & $ 0.296\pm0.085$ & $29$ & $ 0.293\pm 0.126$ & $ 9$ & $ 0.257\pm 0.085$ & $17$ & $ 0.297\pm 0.085$ & $20$ \\
And II  & $-1.365\pm0.117$ & $248$ & $ 0.033\pm0.085$ & $56$ & $ 0.096\pm 0.085$ & $21$ & $ 0.041\pm 0.085$ & $40$ & $-0.049\pm 0.085$ & $33$ \\
And I & $-1.109\pm0.117$ & $ 31$ & $ 0.278\pm0.164$ & $ 7$ &  &  & $ 0.427\pm 0.157$ & $ 5$ & $ 0.192\pm 0.225$ & $ 5$ \\
And III & $-1.808\pm0.117$ & $ 35$ & $ 0.205\pm0.159$ & $ 8$ & $ 0.333\pm 0.208$ & $ 5$ & $ 0.026\pm 0.197$ & $ 5$ & $-0.009\pm 0.312$ & $ 3$ \\
And V & $-1.712\pm0.117$ & $ 80$ & $ 0.119\pm0.085$ & $40$ & $ 0.131\pm 0.085$ & $23$ & $-0.004\pm 0.085$ & $28$ & $-0.014\pm 0.085$ & $14$ \\
And X & $-2.298\pm0.123$ & $ 16$ & $ 0.505\pm0.240$ & $ 5$$^{e}$ & $ 0.368\pm 0.446$ & $ 3$ &  &  &  & 
\enddata
\renewcommand{\thefootnote}{\alph{footnote}}
\footnotetext[1]{\fehavg\, calculated from spectra with S/N $>$ 8 \AA$^{-1}$.}
\footnotetext[2]{\alphafeavg\, calculated from spectra with S/N $>$ 15 \AA$^{-1}$, for the metallicity ranges specified below.}
\footnotetext[3]{\fehavg\, for M~32, NGC~185 and NGC~147 may be biased (low) estimates, as there may be $\feh>0$ stars not measurable by our grid (see \S~\ref{ssec_analysis})}
\footnotetext[4]{\alphafeavg\, for M~32 is not representative of this galaxy's population.}
\footnotetext[5]{\alphafeavg\, for And~X is possibly a biased (high) estimate, since 3 of 5 stars have large $\sigma^{-}$ uncertainties in \alphafe\, (see Figure~\ref{fig: alpha_grid}).}
\label{table: averages}
\renewcommand{\thefootnote}{\arabic{footnote}}
\end{deluxetable*}

\subsubsection{\alphafeavg\,\lowercase{vs} D\lowercase{istance from} H\lowercase{ost}} 
We explore whether the environment plays a role in driving the
variation in chemical properties by plotting the average alpha 
abundance against galactocentric distance from M\,31 (top-right 
panel of Figure~\ref{fig: alphafe_properties}). For 
reference, the plot also includes the same set of comparison
Milky Way dSphs, plotted against galactocentric distance from 
the Milky Way. We do not measure a significant correlation 
with environment, even at a host galaxy separation as large as 
that of And~VII, the most distant M\,31 satellite in our sample. 
If star formation in these dwarfs were very strongly influenced 
by environment, we might expect the objects closer to M\,31 to 
show signs of truncated star formation histories, and thus higher
\alphafeavg. However, one of the most alpha-enhanced systems is 
also one of the most distant (And~VII).  A full analysis of the role 
of environment requires knowledge of the orbital history of each
satellite, which is not currently available. 

\subsubsection{\alphafeavg\,\lowercase{vs} K\lowercase{inematics}} 
We next compare the average alpha abundances against global
galaxy kinematics (bottom-left panel of Figure~\ref{fig: alphafe_properties}).
We plot \alphafeavg\, against $v_{rms}$, defined as the quadrature 
sum of $\sigma_{los}$ and the maximum rotation velocity $v_{rot}$, 
for those galaxies that show significant rotation, and equal to $\sigma_{los}$ 
otherwise.  We use the spatially averaged (not central) $\sigma_{los}$, 
for NGC~185 and NGC~147, since a significant fraction 
of those data extend beyond their respective half-light radius. 
Taking into account both rotation and dispersion, there is no clear 
trend of alpha abundances with kinematics. Claims of such 
correlations have been long reported in extra-galactic studies of 
giant ellipticals \citep[e.g.][]{Trager2000a,Thomas2005a,Conroy2014a,
Greene2013a}. The lack of kinematic-abundance correlation suggests 
a difference in the formation mechanisms of giant versus dwarf elliptical 
(and spheroidal) systems.

\subsubsection{\alphafeavg\,\lowercase{vs} S\lowercase{tellar} D\lowercase{ensity}} 
In the bottom-right panel of Figure~\ref{fig: alphafe_properties} we 
compare the average alpha abundance 
against the central surface brightness, $\mu_{V}$, a proxy for stellar mass 
density, since it could be expected that dwarf galaxies with higher stellar 
density could form stars more efficiently \citep{Revaz2012a}.
We do not detect any clear correlation. However, we note that 
our sample's radial coverage precludes a direct comparison 
at similar radii for all galaxies. The quoted $\mu_{V}$ only 
encapsulates the density in the central region, whereas a 
significant fraction of our stellar sample is beyond the inner 
half-light radius. 

\subsection{Alpha Abundance Trends with Metallicity}\label{ssec_trend_feh}

The distribution of \alphafe\, as a function of \feh\, reflects
the chemical enrichment pathway within a galaxy. Roughly,
the ISM evolves from metal-poor and alpha-enhanced towards 
higher metallicities and lower alpha abundances. While this 
picture is likely overly simplified for complex stellar populations, 
it is a good first step in finding differences between galaxies.
Figure~\ref{fig: alpha_grid} hints at significant differences 
between the various galaxies in our sample. And~V is 
the most salient example for an internal trend of decreasing 
alpha abundance with rising iron abundance, and for a bulk 
population with \mbox{\alphafe$\lesssim{0.0}$}. In contrast, 
And~VII is distinguished by a lack of a trend towards lower alphas: 
the majority of And~VII stars in our sample have \alphafe$\sim{+0.3}$,
and the few outliers with low \alphafe\, do not constitute a trend 
of alpha abundance with metallicity.

\begin{figure*}[tpb!]
\centering
\includegraphics[width=\textwidth]{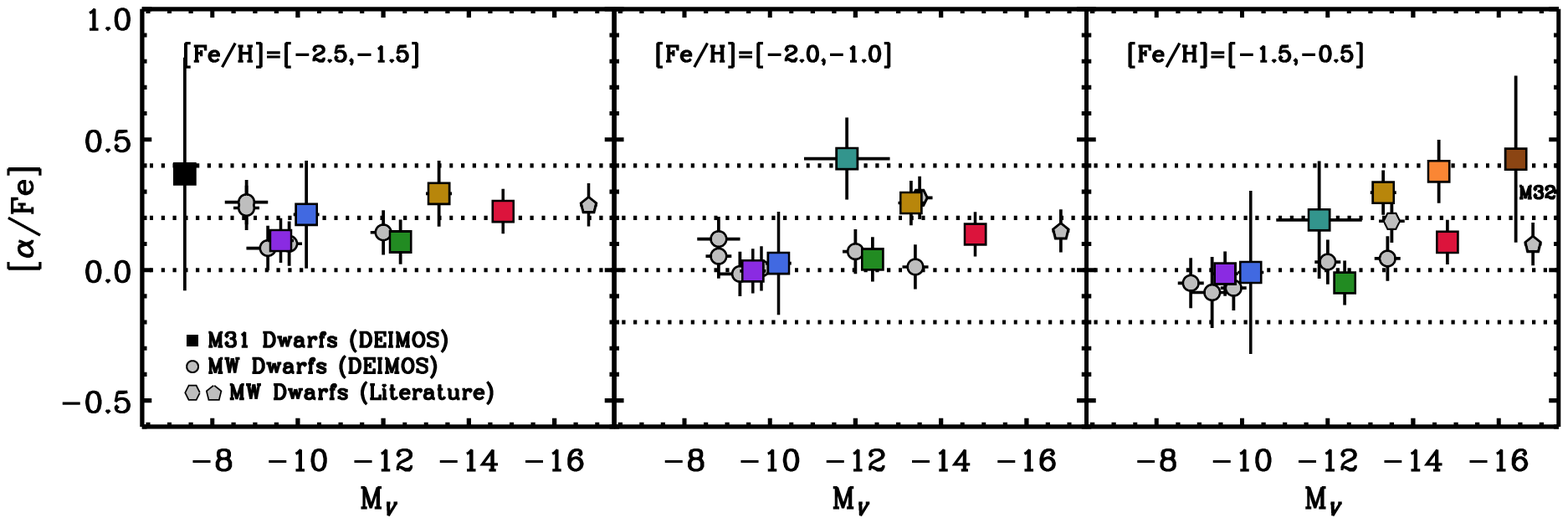}	
\caption{Variation of mean alpha abundance, \alphafeavg. with 
luminosity as a function of \feh.  We plot \alphafeavg\, calculated 
for three partially overlapping metallicity bins containing the 
majority of the stars in the sample. In each panel, only those 
dwarf galaxies with three or more \alphafe\, measurements are
shown in each panel. The color coding is the same as in 
Figure~\ref{fig: alphafe_properties}. The comparison Milky Way 
(MW) dwarf satellite data are shown as gray symbols. Gray circles 
denote MW dSphs from DEIMOS data, reanalyzed with our code. 
We also add literature abundances for the Sagittarius dSph 
(gray pentagons) from \citet{Carretta2010b} and the Small 
Magellanic Cloud (gray hexagons) from \citet{Mucciarelli2014a}.
The dotted horizontal lines are added for reference in order 
to better distinguish changes in alpha abundance across the 
three metallicity bins. }
\label{fig: alphafe_luminosity}
\end{figure*}

We seek to better quantify these trends, accounting for the 
different sample sizes.  We thus choose three partially overlapping 
metallicity bins, $-2.5<\feh<-1.5$, $-2.0<\feh<-1.0$, $-1.5<\feh<-0.5$. 
We do not include higher or lower metallicities due to 
the lack of a correction function for \alphafeatm, and a 
paucity of measured abundances, respectively. For each 
metallicity bin, we measure the average alpha abundance 
ratio in each M\,31 dwarf galaxy, as well as the MW 
comparison sample, as described in \S~\ref{ssec_meantrends}.  
For this comparison, we also add literature abundances for 
the Sagittarius dSph (gray pentagons) from \citet{Carretta2010b} 
and the Small Magellanic Cloud (gray hexagons) from 
\citet[][priv. comm.]{Mucciarelli2014a}. Not all dwarf galaxies 
populate each bin, due to the different iron abundance ranges 
spanned by each system. Figure~\ref{fig: alphafe_luminosity} 
plots the resulting \alphafeavg\, trend with luminosity for each metallicity 
bin, only for galaxies with three or more stars in each bin.
Four out of five M\,31 satellites appearing in all three bins 
show a gentle decrease in \alphafeavg\, for increasingly 
higher metallicity bins (And~II, And~III, And~V, and NGC~185).

In the most metal-poor bin (left panel), \alphafeavg\, is 
indistinguishable between the four MW dwarf galaxies
(Sextans, Leo~II, Leo~I, and Fornax) and three M\,31 
dwarf galaxies (And~V, And~III, And~II) that overlap in 
absolute magnitude. The two brighter M\,31 satellites in 
the figure, And~VII and NGC~185, are slightly more 
enhanced by $\sim{0.1-0.2}$ dex, and so is the SMC
datapoint. In the intermediate metallicity bin, the same 
three M\,31 satellites, And~V, And~III, and And~II, have 
\alphafeavg\, consistent with that of Sextans, Leo~II, 
Leo~I, and Fornax.  And~VII and And~I have enhanced 
alpha abundance ratios, similar to the Sag dSph 
(\alphafeavg$\sim{+0.3}$). NGC~185 and the SMC 
have \alphafeavg$\sim{+0.2}$.

We detect a larger variation in \alphafeavg\, at the highest 
metallicity bin. In this bin, And~V, And~III, And~II, and 
NGC~185 have \alphafeavg\, consistent with the MW satellites, 
However, the alpha abundances for And~VII and 
NGC~147 (and perhaps M\,32 but with large uncertainties) 
are enhanced relative to the MW dwarf galaxies at similar 
luminosity. Thus, in contrast to the MW satellite population, 
the M\,31 dwarf galaxies exhibit a larger variation in alpha 
enhancements, in particular for M$_{V} < -12$ and
$\feh\gtrsim{-1.5}$. We strongly caution that these trends 
should not be extrapolated to fainter magnitudes.

\begin{deluxetable*}{lrrrrrr}[t!]
\tablecolumns{7}
\tabletypesize{\footnotesize}
\tablecaption{Chemical Abundance Radial Gradients}
\tablewidth{0pt}
\tablehead{
\colhead{Galaxy} & 
\colhead{d[Fe/H]/d(r/r$_{1/2}$)$^{a}$} & 
\colhead{P$_{KS}$(\feh)$^{b}$} & 
\colhead{Gradient \feh\,?} & 
\colhead{d[$\alpha$/Fe]/d(r/r$_{1/2}$)$^{a}$} & 
\colhead{P$_{KS}$(\alphafe)$^{b}$} & 
\colhead{Gradient \alphafe\,?} 
}
\startdata
NGC185 & $ -0.085\pm0.004$ & $0.0000916$ & Yes & $  0.033\pm0.012$ & $0.0703$ & Not sig. (Yes$^{d}$)  \\
NGC147 & $  0.015\pm0.005$ & $0.00839$ & Not sig. & $  0.138\pm0.068$ & N/A & Not sig. \\
AndVII & $ -0.169\pm0.037$ & $0.790$ & Not sig. & $  0.082\pm0.080$ & $0.344$ & Not sig. \\
AndII & $ -0.448\pm0.023$ & $0.00000872$ & Yes & $  0.039\pm0.063$ & $0.917$ & Not sig. \\
AndV & $ -0.169\pm0.029$ & $0.479$ & Not sig. & $ -0.026\pm0.064$ & $0.993$ & Not sig.
\enddata
\renewcommand{\thefootnote}{\alph{footnote}}
\footnotetext[1]{Best-fit linear slope and its 1$-\sigma$ uncertainty.}
\footnotetext[2]{K-S probability that the inner-half and outer-half subsamples are drawn from same population.}
\footnotetext[3]{Criteria for significance of gradient: (i) Slope has to be $3\sigma$ away from zero, and (ii) P$_{KS} < 0.05$.}
\footnotetext[4]{Gradient is significant \textit{if} $\feh>-0.5$ stars are included.  This is driven by a few \alphafe<0 stars,
see also Figure~\ref{fig: abundances_radius}.}
\label{table: gradients}
\renewcommand{\thefootnote}{\arabic{footnote}}
\end{deluxetable*}

\subsection{Radial Abundance Gradients}\label{ssec_radial}

Radial abundance gradients, if present, may be indicators 
of internal variation in the chemical evolutionary history of 
dwarf galaxies.  We restrict our analysis of iron and alpha 
gradients to the five galaxies with the largest samples (NGC~185, 
NGC~147, And~II, And~VII, and And~V).  We test for the 
presence of radial abundance gradients in two ways. First, 
we fit linear functions to \feh\, and \alphafe\, as a function
of (projected) elliptical radius. The radii are scaled to the 
half-light radius of each galaxy, shown in Table~\ref{table: properties}. 
To consider a gradient significant, we require that the linear slope
deviate from zero by \textit{at least} $3\sigma$. We then 
split each galaxy's \feh\, and \alphafe\, samples in two radial bins, 
each subpopulation containing 50\% of the stars (ordered 
by projected distance to the galaxy's center). Using the 
Kolmogorov-Smirnov (K-S) statistic, we ask 
what is the probability that the inner/outer \feh\, and \alphafe\, 
subsamples are inconsistent with being drawn from the same 
population with 95\% confidence, i.e. P$_{KS} < 0.05$. We only
regard as significant those gradients that pass both tests 
above. Given our spatial sampling, we note that the samples 
for And~II, And~V, and  And~VII extend roughly out to two 
half-light radii, while our sampling for NGC~185 and NGC~147
extends to many half-light radii. 

\begin{figure*}[tpb!]
\centering
\includegraphics[width=\textwidth]{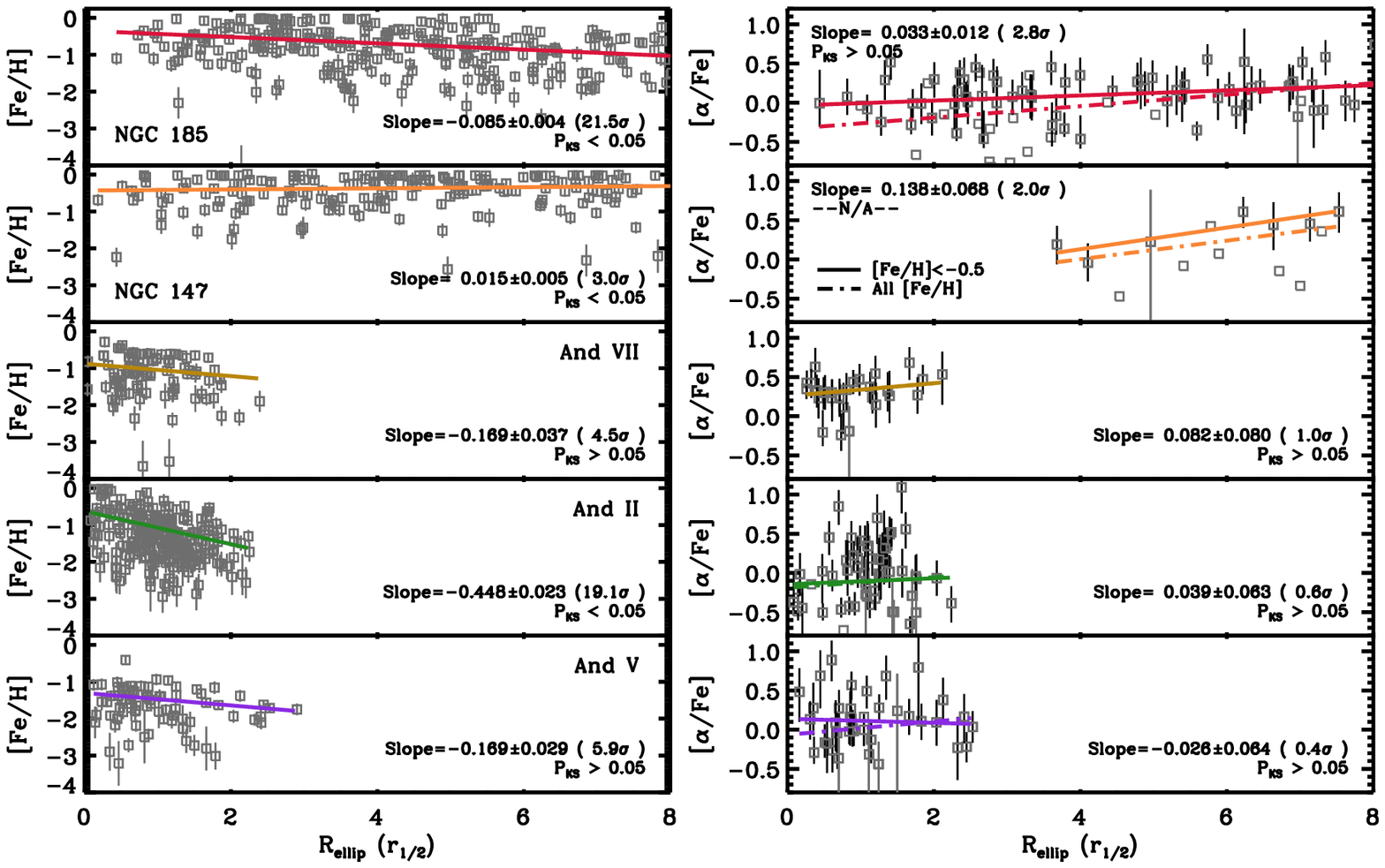}	
\caption{Radial gradients in iron and alpha abundance for the five
dwarf galaxies with the largest sample sizes, shown as empty squares with
$1\sigma$ error bars. The \feh\, panels show all data with \mbox{S/N$\geq{8}$\AA$^{-1}$}, 
whereas the \alphafe\, panels show only data with \mbox{S/N$\geq{15}$\AA$^{-1}$} 
(the \alphafe\, sample).  In the \alphafe\, panels, squares with error bars
indicate stars with $\feh<-0.5$, whereas more metal-rich objects are 
plotted without uncertainties.  The best-fit \feh\,  radial gradients are 
plotted as solid lines,  and the best-fit values are reported in each panel.
The \alphafe\, gradient fits calculated using only \feh$<-0.5$ data are 
shown as solid lines, and the slopes are also included in each panel.  
We also plot as dot-dashed lines the \alphafe\, gradient fits 
including the $\feh>-0.5$ data, but caution that these \alphafe\, abundances 
may not be reliable, as discussed in \S~\ref{ssec_analysis}.
The abundances are plotted against elliptical radius. The radii have been rescaled in units 
of each dwarf galaxy's half-light radius, $r_{1/2}$, shown in Table~\ref{table: properties}. 
The half-light radii and gradient slopes and 1$-\sigma$ uncertainties are plotted on each 
panel. The gradients are tabulated in Table~\ref{table: gradients}.}
\label{fig: abundances_radius}
\end{figure*}

We plot the \feh\, and \alphafe\, samples as a function of  radius in
Figure~\ref{fig: abundances_radius}.  The best-fit slopes, as well as
the probability of the samples being drawn from the same population,
P$_{KS}$, are shown in Table~\ref{table: gradients}.  On the basis of
the two criteria described above, we detect a significant \alphafe\, radial 
gradient only in NGC~185. One caveat is that our detection is sensitive 
to the inclusion of the small fraction of \feh$>-0.5$ stars suffering from 
a lack of well-tested \alphafe\,-\alphafeatm\, calibration (see \S~\ref{ssec_analysis});
if these stars are excluded, the gradient signature is no longer significant.
And~VII, And~V, and And~II do not show any evidence for an alpha 
abundance gradient via either test.  We do not apply the K-S test to 
the NGC~147 \alphafe\, sample because of the limited radial coverage 
of the \alphafe\, measurements. However, the slope in the outer regions
is less than $3-\sigma$ away from zero. The lack of detectable \alphafe\, 
radial gradients suggests that the star formation timescales were not 
strongly dependent on the radial distance from the center of the galaxy.  
In the case of And~VII, high \alphafe\, abundances dominate at all radii, 
suggestive of a global short star formation timescale (the few low 
\alphafe\, stars are discussed in \S~\ref{ssec_disc_evolution}). 
In contrast,  And~V and and~II have both low and high \alphafe\, values at all 
radii, indicative of formation timescale long enough for a significant influence 
by Type~Ia SNe. Out of the samples for six comparison MW dSphs, four 
extend beyond one half-light radius (Leo~I, Leo~II, Draco, and Sextans). 
We do not detec significant \alphafe\, gradients in any of them. 

We next consider the presence of \feh\, radial gradients.  We 
detect statistically significant radial gradients in NGC~185 and
And~II.  Both galaxies have gradient slopes $>3\sigma$ different 
from zero, and P$_{KS} << 0.05$.  We do not detect significant iron 
abundance gradients in the other three galaxies,  And~VII,  And~V 
and NGC~147.  Our results our consistent with the findings by 
\citet{Ho2014a} for the four galaxies in common with their sample 
(And~V is not included).  Among the four comparison MW dSphs 
with radial coverage larger than one half-light radius, we measure 
significant \feh\, radial gradients in Leo~I and Leo~II.  The presence 
of iron abundance gradients may be an indication of inside-out growth, 
radial outflows or inflows \citep[e.g.,][]{Schroyen2013a,Revaz2012a}. 

With the possible exception of NGC~185 (see caveat above), 
our combined M\,31 and MW sample is consistent with a lack
of discernible \alphafe\, radial gradients. In contrast, metallicity
gradients are present in some of the dwarf galaxies of both the
MW and M\,31. In the absence of a significant alpha abundance 
gradient, a metallicity gradient can still be interpreted as a difference 
in star formation efficiency, such that the outer, less bound regions, 
had greater difficulty holding on to their gas component \citep[e.g.][]{Lanfranchi2004a},
but both inner and outer regions evolved for long enough to
have high and low \alphafe\, stars. In the two M\,31 satellites lacking
both iron and alpha abundance gradients, the 
ISM may have mixed efficiently over the star formation period of 
the galaxy \citep{Mac-Low1999a}.  We end with the caveat that
gradients may not be visible unless data extends for many half-light
radii, and our results hold for the spatial range covered by our 
samples. The case of Fornax in the comparison MW sample is 
instructive; while no gradient is detected in the central region targeted by 
\citet{Kirby2010a}, more spatially extended datasets could 
measure hitherto undetected abundance gradients 
\citep[e.g.,][]{Leaman2013a}.

\section{Discussion}\label{sec_disc}

\subsection{Evolutionary Constraints on the M\,31 Dwarf Galaxies}\label{ssec_disc_evolution}

The combination of alpha and iron abundances serves as a star 
formation clock due to the different timescales between enrichment
from Type~II and Type~Ia SNe ejecta. The M\,31 system contains 
satellite galaxies with a range of average alpha abundances from 
$\alphafeavg\sim{+0.0}$ to $\alphafeavg\sim{+0.5}$ (see 
Table~\ref{table: averages}).  This implies a range of star formation 
timescales among the M\,31 dwarf galaxies. The bulk of stars in 
dwarf galaxies with high \alphafe\, formed in short timescales in 
ISM environments rich in Type~II SNe ejecta, whereas galaxies 
with low \alphafe\, tended to form their stars in an ISM which 
was also polluted by Type~Ia SNe operating on longer timescales.  

We discuss the star formation timescales of M\,31 satellites implied 
from this work in combination with literature results on their stellar 
ages. The fainter six dwarfs galaxies were characterized as having
primarily "old" stars formed $\gtrsim{10}$ Gyr ago based on their 
horizontal branch populations \citep{Grebel1999a, Da-Costa2002a}, 
with the possible exception of And~II \citep{Da-Costa2000a,McConnachie2007a}. 
Significant intermediate age populations are excluded by the absence 
of large numbers of asymptotic giant branch (AGB) stars \citep[e.g.,][]{Armandroff1993a,Grebel1999a}. 
The recent star formation histories (SFHs) by \citet{Weisz2014a} suggest 
that at least 50\% of the stellar mass of And~I, And~III, and And~V 
was in place more than 10 Gyr ago, with the remaining stellar mass 
forming between $\sim{10}$ and 7 Gyr ago (their Figure 7), but 
systematic uncertainties in the oldest age bins may lower this fraction 
of younger stars. In contrast, the evidence
for an intermediate age stars has been confirmed by \citet{Weisz2014c},
and is robust to systematic uncertainties. We discuss And~II  
separately in later paragraphs due to its difference in stellar age
distribution. In spite of their old(er) stellar ages, the faintest galaxies
in our sample exhibit a variety of patterns in \alphafe-\feh\, 
space. The majority have both low and high alpha abundance stars. 
The combination of old ages and decreasing \alphafe\, with rising \feh\, 
resembles the abundance patterns of some of the MW classical dSphs 
with old stellar populations, such as Sculptor \citep{Tolstoy2009a,Kirby2011b}. 
In the case of Sculptor, \citet{De-Boer2012b} interprets this type of chemical 
abundance patterns as consistent with a slow burst of star formation, 
tapering off over a few Gyrs, and it appears that the influence of 
Type~Ia SNe takes longer to become evident on the chemical 
abundances than could be naively assumed by taking the minimum 
age for prompt Type~Ia SNe, $\sim{100}$ Myr \citep{Maoz2012a}. Given 
their age and abundance similarity, the chemical evolution of the
faint M\,31 dwarf galaxies, excluding And~VII and And~II, 
may be similar to that of the Sculptor dSph.

Among the fainter satellites,  And~VII has a clearly elevated characteristic 
alpha abundance, suggesting a short star formation timescale. The lack 
of a decrease in alpha abundances with rising metallicity suggests that 
star formation did not last long enough for Type~Ia SNe to significantly 
enrich the ISM. Star formation may have been truncated by a close 
encounter with M\,31 \citep[e.g.,][]{Nichols2014a} before Type~Ia SNe 
had time to influence the alpha abundances. Given its present large 
separation from M\,31 today (> 200 kpc, \citealt{McConnachie2012a}), 
this would suggest a closer separation to M\,31 in the past. Recently, 
\citet{Weisz2014a} found a hint of such a short SFH in their analysis of 
HST/WFPC2 data, but their And~VII photometry was too shallow. 
Deeper photometry should place interesting constraints on its SFH. 

In spite of And~VII's overall chemical abundance trend, a few low alpha 
abundance outliers are present at low metallicities.  
The occurrence of these low \alphafe\, stars could be the explained 
if And~VII's abundance pattern is the sum of more than one stellar population
each with different star formation characteristics.  The low alpha, low \feh\, 
stars could then be explained as coming from the secondary stellar population,
e.g. from satellite-satellite accretion, or from a dissolved star cluster \citep[e.g.,][]{Karlsson2012a}.
If so,  the secondary system would have to be a smaller population in order 
not to dominate the overall $\alphafe-\feh$ abundance pattern. Example of 
anomalous abundance patterns are also found in other environments, such 
as the Milky Way halo \citep[e.g.,][]{Ivans2003a}, and are interpreted similarly 
but in the context of halo formation.  An alternate explanation is that star 
formation was dominated by bursts, leading to repeated enhancement in 
alpha from Type~II SNe \citep{Gilmore1991a}. The presence of these low 
alpha stars would then be consistent with stars formed in the time 
between bursts, when Type~Ia SNe acted to decrease \alphafe.

\begin{figure*}[tpb!]
\centering
\includegraphics[width=\textwidth]{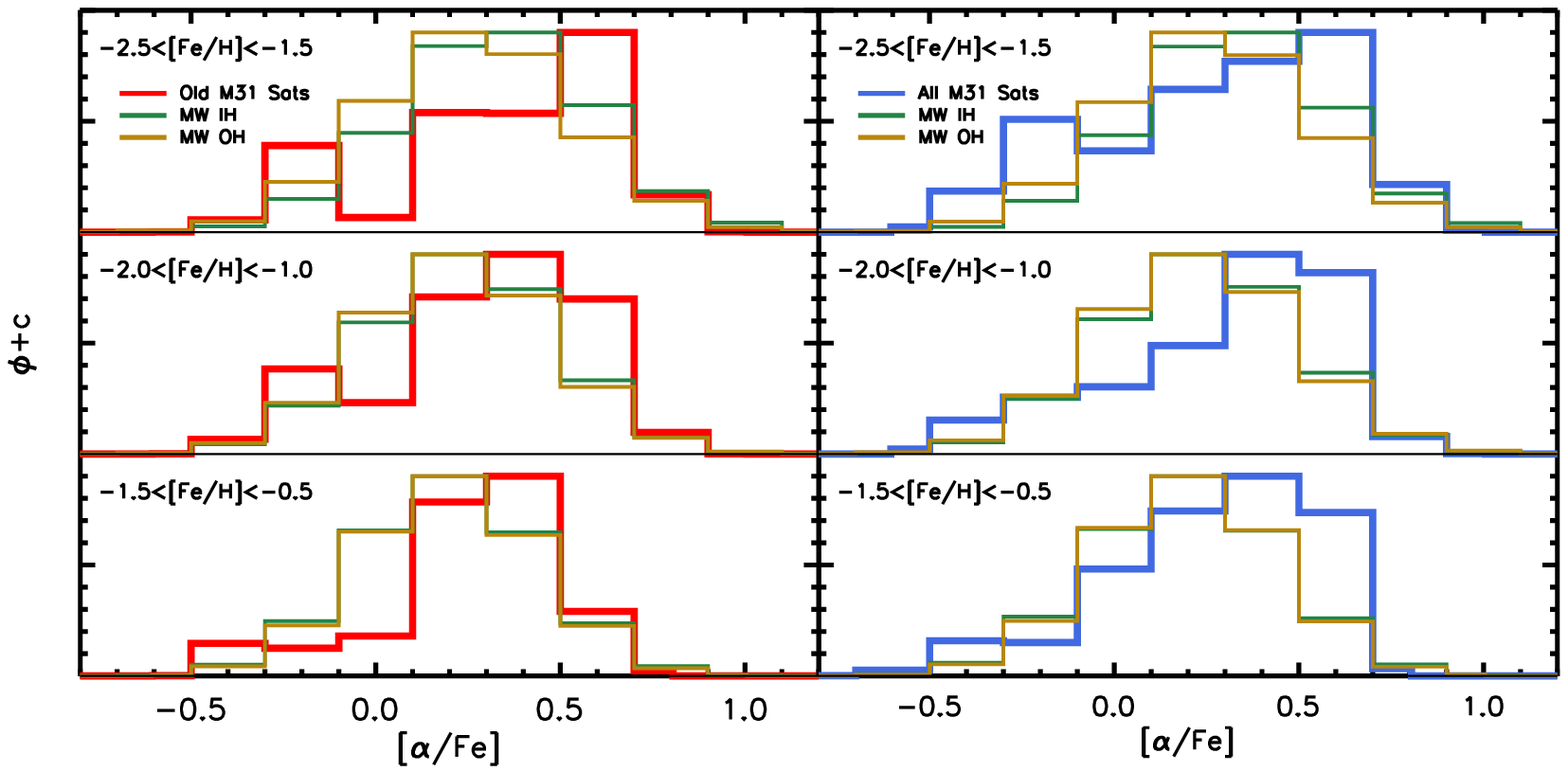}	
\caption{Comparison of the (normalized) \alphafe\, distribution
of a hypothetical stellar halo made up of M\,31 satellites, as described 
in \S~\ref{ssec_disc_halo}, against the Milky Way (MW) halo sample 
of \citet{Ishigaki2012a}. The M\,31 \alphafe\, distribution in the left
panels (thick red histograms) includes only M\,31 satellites with old stellar 
populations, whereas the right panels includes all satellites (thick blue 
histograms). The MW halo sample is split into kinematically-selected 
inner (thin green) and outer (thin gold) halo components. From top 
to bottom, we include only stars in the three metallic bins denoted in 
each panel. The distribution of \alphafe\, for the M\,31 satellite samples 
disagrees with that of the MW samples both by having a lower mean 
value, and a more skewed distribution, Thus, that a stellar halo made 
up of present day M\,31 satellites could not form a MW-like halo, 
even if including only dwarf satellites with dominant old stellar 
populations.}
\label{fig: halo_afe}
\end{figure*}

In contrast to the previously discussed satellites, And~II and 
the brighter systems (NGC~185, NGC~147, and M\,32) built 
their stellar populations more gradually, and reached $\sim{90}$\%
of their total mass by $\sim{5}$ Gyr ago in the case of And~II \citep{Weisz2014c}
and 2-4 Gyr ago for the others \citep{Weisz2014a}. Their CMD 
morphologies also feature red clumps, AGB populations, 
and in some cases carbon stars \citep[e.g.,][]{Butler2005a,Battinelli2004b,McConnachie2007a}. 
Both And~II\footnote{Recently, \citet{Amorisco2014a}
postulated the presence of a stream in And~II confined to a 
region between R$_{\rm{And~II}}$$\sim{0.9}$ and $\sim{1.7}$ 
kpc (projected), excluding a small range of projected angles. 
We have compared the distribution of stellar \alphafe\,and \feh\, 
between their high-likelihood stream and non-stream members
(\textit{priv. comm.}) but do not detect a difference in the chemistries.} 
and NGC~185 have low average alpha abundances at a large range of 
metallicities. NGC~147, while having a high average alpha abundance 
for \mbox{\feh$<-0.5$}, may reach low alpha abundances  for its highest 
metallicity stars. This pattern may imply that an abundance "knee" in NGC~147, 
indicative of the onset of Type~Ia enrichment appears at a very 
high metallicity, suggesting a very efficient initial chemical enrichment.  
However, the alpha abundances for \mbox{$\feh>-0.5$} are uncertain 
due to the lack of a calibrator sample for our \alphafe\, correction term 
(\S~\ref{ssec_analysis}). The presence of many stars with low alpha 
abundances in And~II, NGC~185, and NGC~147 is consistent with a 
long lasting star formation process. In addition,  And~II and NGC~185 
show large scatter in their \alphafe-\feh\, plane at fixed \feh.
This may simply reflect some level of inhomogeneous chemical 
enrichment, such that not all ISM regions are polluted with the 
same amounts of Type~II and Type~Ia products.  An alternate 
way to explain this scatter may be to postulate the presence of 
more than one track in \alphafe\,-\feh\, space, thus boosting the
scatter at fixed metallicity. If so, these populations should be 
mixed radially, in order to be consistent with the lack of radial alpha abundance 
gradients discussed in \S~\ref{ssec_radial}. One way to probe 
this is by checking whether there are age differences between 
the high and low alpha abundance stars at a fixed metallicity. 

\subsection{The Present Day M\,31 Dwarfs and the Build-up of M\,31's Halo}\label{ssec_disc_halo}

In $\Lambda$CDM, stellar halos are expected to form 
at least in part through accretion of smaller systems
\citep[e.g.,][]{Searle1978a,Bullock2005a,Bell2008a,
Cooper2010a,Font2011b}. The chemical abundance 
patterns of these accreted systems thus survive as 
imprints in the present day stellar halo. While the
metallicity of the M\,31 halo has been the subject
of recent work \citep[e.g.,][]{Reitzel2002a,Gilbert2006a,Koch2008a},
the only stellar halo for which \alphafe\, measurements 
are available is that of the MW, so we limit our 
comparisons to this population. The distribution of 
chemical abundances in MW inner halo stars disagrees 
with that of the present day MW dSphs \citep[][V13]{Venn2004a}.
The inner halo stars likely formed earlier than 10~Gyr 
ago \citep[e.g.,][]{Jofre2011a,Schuster2012a,Kalirai2012a}. 
Motivated by the differences in the M\,31 dwarf satellite 
population relative to the MW satellites, and the fact that 
some of the M\,31 satellites appear to be have old stellar 
populations systems, we now ask what type of halo would 
be formed out of the present day M\,31 dwarf galaxies, 
and how similar such a halo would be to what is seen in 
the MW. 

\begin{figure}[tpb!]
\centering
\includegraphics[width=.48\textwidth]{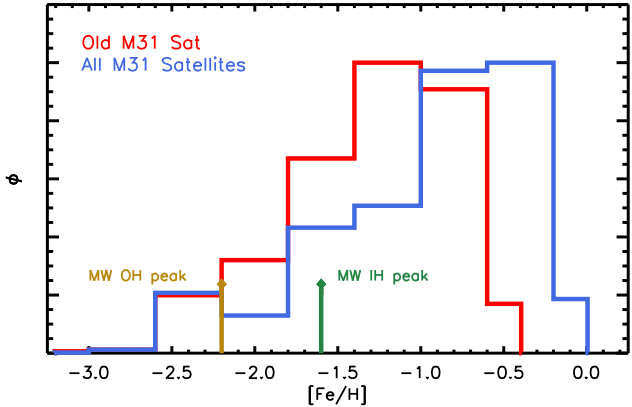}	
\caption{Normalized metallicity distributions for a stellar 
halo made up of destroyed M\,31 satellites (same colors
as in Figure~\ref{fig: halo_afe}). The arrows indicate the 
peaks of the metallicity distributions for the Milky Way 
inner and outer halos, from \citet{Carollo2010a}. A halo 
made up of present-day M\,31 satellites would be too 
iron-rich in contrast to either MW stellar halo population, 
even if only considering the M\,31 dwarf satellites with 
a majority old population.}
\label{fig: halo_feh}
\end{figure}

We thus combine our dwarf galaxy samples into a single \alphafe-\feh\, 
distribution. However, simply looking at the raw distributions of stars 
in our sample is incorrect. This is because the variable crowding 
of stars and limited size of our slitmasks prevents a uniform sampling 
of the M\.31 satellites.  Hence, we weight each star to take into account 
the the stellar mass of each galaxy ($M_{*,j}$), and the luminosity function 
of M\,31 galaxies. We assign a weight to each star in our sample for 
galaxy \textit{j}, using Equation~\ref{eqn: weights}, where the index $j$ 
represents the sample's galaxies ordered by luminosity. 

\begin{equation}
w_{i,j} = \frac{M_{*,j}[\Phi{}(L > 0.5(L_{j}+L_{j-1}))-\Phi{}(L > 0.5(L_{j}+L_{j+1})]}{N_{j}}
\label{eqn: weights}
\end{equation}

Here, \mbox{$\Phi{}(L>L_{x})$} is the cumulative 
luminosity function of M\,31 dwarf galaxies brighter than 
a given luminosity, and $N_{j}$ is the size of the sample for 
galaxy \textit{j}. For the M\,31 dwarf satellite luminosity 
function, we assume a simple functional form with two 
broken linear slopes, which roughly approximates the trend 
presented in the left panel of Figure~1 by \citet{Yniguez2014a}. 
We then use the weights \textit{w} to create \textit{w} samples of 
each star, each with the same \feh\, and \alphafe\, values as the 
original stars. This  corrects for our non-uniform sampling 
of each dwarf galaxy's stars, as well as the fact that we have 
a larger fraction of the bright M31 satellites in our sample, 
but only a few of the faint ones. We have excluded M\,32
from all following comparisons because we lack a representative 
sample (only five stars), and because it is a very unusual object 
\citep{Howley2013a}, so that its contribution to a stellar halo 
may not be important in most cases.

We compare the distribution of alpha abundances between the 
combined M\,31 satellite sample and the present day MW halo 
as a function of metallicity. We choose the same metallicity 
bins as in Figure~\ref{fig: alphafe_luminosity}. For the MW halo comparison
sample, we use the very high-resolution data by \citet{Ishigaki2012a}, 
who subdivide their population into inner and outer halo populations 
based on stellar kinematics. We only use stars in that sample for 
which all of [Mg/Fe], [Si/Fe], [Ca/Fe], and [\ion{Ti}{1}/Fe] are measured 
\footnotetext{We use \ion{Ti}{1} because our spectra also measure 
primarily neutral Ti lines.}. We define \alphafe\, for the MW sample 
as the unweighted mean of all four abundance ratios, consistent with 
the definition for our sample (see \S~\ref{ssec_analysis}). The typical
uncertainty in \alphafe\, for the halo stars, $\sim{0.05}$ dex, is smaller
than for our M\,31 measurements. For a more appropriate M\,31-MW
comparison, we add Gaussian noise to each individual MW halo \alphafe\, 
value, such that their typical \alphafe\, uncertainty is then comparable 
to that of M\,31 satellite stars. We create $\sim{1000}$ realizations of 
each MW halo \alphafe\, measurement using different random seeds. 

Figure~\ref{fig: halo_afe} shows histograms of the distribution in \alphafe\, 
for the three metallicity bins that encompass most of our data and that of 
the MW stars (the same bins as in Figure~\ref{fig: alphafe_luminosity}). 
All histograms have been normalized to their peak value. For all three bins, 
both the MW inner and outer halo and M\,31 satellite samples are 
alpha-enhanced. The peak of the M\,31 samples are higher than the 
MW halo samples by $\sim{0.1-0.3}$ dex, with the difference 
being largest in the lowest metallicity bin. Differences in \alphafe\, between 
the inner/outer MW halo have been reported by \citet{Roederer2009a} and
\citet{Ishigaki2012a}, but as seen from the plot, are too small to be 
measurable when convolving the MW data with our typical data uncertainties.
Nevertheless, we consider the inner and outer MW halo samples separately. 

We formally compare the \alphafe\, distributions by performing K-S tests 
between (a) the M\,31 satellite sample from old satellites only, (b) all satellites 
including those with intermediate age satellites), against the MW halo samples, 
(1) inner halo, and (2) outer halo. We perform this comparison at each metallicity
bin, resulting in twelve comparison pairs. We use the associated P$_{KS}$ values
to assess whether the samples are inconsistent with being drawn from the same 
distribution. For each comparison, we recalculate P$_{KS}$ by randomly sampling 
400 subpopulations each with the same number of stars as the raw data samples 
(Table~\ref{table: abundances}). This results in 400 estimates of P$_{KS}$ for 
each comparison. In every comparison of MW halo versus all (i.e., old plus intermediate age) 
M\,31 satellites, the P$_{KS}$ values are less than $10^{-4}$. For the comparisons 
with old-only M\,31 satellites, the majority of realizations of P$_{KS}$ peak at $< 0.1$, 
in all but two of the comparisons. We conclude that the data is inconsistent with 
the MW halo and M\,31 dwarf satellite samples being drawn from the 
same parent distribution with more than 90\% confidence, except for 
small regions in $\alphafe-\feh\,$ space. The overall difference in 
\alphafe\, suggests that a halo made up of destroyed M\,31 dwarf satellites 
(or satellites sharing their chemical evolution properties) could not have 
been the primary contributor to a MW-like halo. 

We next compare the metallicity distribution from M\,31 
satellites to that of the MW halo. Figure~\ref{fig: halo_feh} 
plots the metallicity distributions. The red histogram
plots the subsample of  dwarf galaxies thought to be mainly old 
(And~I, And~III, And~V, And~VII). The blue histogram also 
incorporates And~II, NGC~185 and NGC ~147, which have moderate 
to strong evidence for significant intermediate age populations (see brief 
age discussion in \S~\ref{ssec_disc_evolution}).  The figure clearly shows 
from the different histogram peaks that a halo made up of old M\,31 satellites 
would be more metal-poor than one which includes the dwarf galaxies with 
intermediate age populations. This is a reflection of the metallicity-luminosity 
relation \citep{Kirby2013b}, since the satellites in our sample with intermediate populations also 
tend to be the most luminous. We compare the peaks of our histograms to 
the peak metallicities of the MW inner and outer halo from \citet{Carollo2010a} 
as vertical lines at $\feh = -1.6$ and $\feh = -2.2$, respectively. The discrepancy 
of peak iron abundances strongly implies that a halo made up of satellites like 
those of M\,31 would result in a halo population too metal-rich in comparison
to the MW halo, regardless of the exclusion of satellites with intermediate
age populations. 

The comparison of alpha abundances and metallicity distributions
shows that a MW halo could not be formed from a population
of present-day M\,31 satellites, even if we only consider those satellites
with predominately old stellar populations. Thus, the old population
dwarf galaxies that formed the MW halo were different from both 
the present-day "old" MW dwarf galaxies and "old" M\,31 satellites,
as has also been suggested in simulations by \citet{Robertson2005a} 
in the context of the MW halo formation. 

\section{Conclusions}\label{sec_concl}

We have measured alpha abundance ratios and stellar metallicities for
\nstars\, individual RGB stars in \nands\, dwarf galaxies of M\,31. We 
first summarize the results internal to our sample:

\begin{enumerate}
\item The stellar populations of the M\,31 dwarf galaxies show 
a variety of "internal" \alphafe-\feh\, abundance patterns, including 
flat trends of \alphafe\, with \feh\, (e.g., And~VII), and decreasing 
\alphafe\, with \feh\, (e.g., And~V).
\item The average alpha abundances ratios vary 
between galaxies from solar \mbox{\alphafeavg$\sim{+0.0}$}, to 
alpha-enhanced \mbox{\alphafeavg$\sim{+0.5}$}, indicating a range 
in star formation timescales. 
\item The average alpha abundance ratios are not significantly
correlated with galaxy luminosity, distance from M\,31 (i.e., environment),
stellar kinematics, or surface brightness. 
\item The variation in alpha abundances between M\,31 dwarf 
galaxies with M$_{V} < -12$ is larger at higher metallicities, 
\feh$\gtrsim{-1.5}$.
 \item We confirm significant radial iron abundance gradients in two 
galaxies out of the five with sufficient data (NGC~185 and  And~II). 
There is only tentative evidence for an alpha abundance radial 
gradient in NGC~185. 
\end{enumerate}

We have homogeneously compared our sample to a sample of 
Milky Way dwarf galaxy satellites, as well as to a literature 
sample of Milky Way inner and outer halo populations. We 
summarize the conclusions from these comparisons below:

\begin{enumerate}
\setcounter{enumi}{5}
\item For \mbox{M$_{V}<-12$}, the M\,31 dwarf satellites show
marginal evidence for larger variation in alpha abundance at
fixed metallicity relative to the Milky Way satellites, especially
at higher metallicities,  \feh$>-1.5$.
\item In the absence of a sample of M\,31 halo alpha abundance, 
we compare the population of M\,31 satellites to the Milky Way 
stellar halo. A stellar halo made from disrupted M\,31 dwarf satellites 
would be more iron-rich \emph{and} have a different alpha 
abundance distribution than the Milky Way halo. This holds 
both for old M\,31 satellites and those with younger stellar 
populations. Therefore, the Milky Way stellar halo could not 
be formed from a population of dwarf galaxies with chemical 
properties like the present-day M\,31 satellites.
\end{enumerate}

The measurements of alpha abundances for M\,31 satellites 
constitute an important contribution to our growing 
understanding of dwarf galaxies. The combination of  accurate 
age distributions and chemical information (e.g., \alphafe$-$\feh\, 
distributions, metallicity distribution functions, radial abundance 
gradients) will help in understanding the detailed evolution in these 
systems, thus expanding galactic archaeology beyond the Milky 
Way. Studies of M\,31 galaxies are also complemented by those 
of nearby field dwarf galaxies at comparable distances. 

With the study of the M31 dwarf galaxies (and systems at similar 
distances), we approach the limit of capabilities for medium$-$resolution 
spectroscopy of individual RGB stars with 8-10 meter class telescopes. 
Within the current telescope constraints, chemical abundance measurements of 
more distant stellar populations are still possible however by co$-$addition 
of spectra from stars with similar stellar parameters, as has been recently 
shown in the case of RGB stars by \citet{Yang2013a}, observations of 
very bright (but relatively young) blue and red supergiants 
\citep[e.g.,][]{Venn2000a,Davies2010a}, and from integrated light 
measurements of stellar populations such as globular clusters \citep[e.g.][]{Colucci2013a}. 
Thirty meter class telescopes now under development will 
greatly expand the spatial region accessible to studies of the 
chemical inventory of nearby galaxies from individual stars. 

\section*{Acknowledgements}

The authors wish to acknowledge extensive help from
Evan Kirby on code development as well as many   
comments that helped improve the science discussion.
We thank Raja Guhathakurta and the SPLASH team 
for helpful conversations and making the present work
possible. We also acknowledge the anonymous referee 
for a useful report that  contributed to improving this 
manuscript, as well as useful discussions with Nhung 
Ho and Ana Bonaca. 

LCV was supported by the National Science Foundation Graduate 
Research Fellowship under Grant No. \mbox{DGE$-$1122492}. LCV and 
MG acknowledge support from NSF Grant \mbox{AST$-$0908752.} 
EJT was supported by NASA through Hubble Fellowship Grant No. 51316.01 
awarded by the Space Telescope Science Institute, which is operated 
by the Association of Universities for Research in Astronomy, Inc., for 
NASA, under contract NAS 5-26555. 

\textit{Facilities:} Keck II (DEIMOS)

\bibliographystyle{apj}

\end{document}